\documentclass[9pt, aps, prd, nofootinbib, twocolumn]{revtex4-2}

\usepackage{amsmath}
\usepackage{amssymb}
\usepackage{bm}
\usepackage{braket}
\usepackage[usenames, dvipsnames]{color}
\usepackage{dcolumn}
\usepackage{diagbox} 
\usepackage{epsfig}
\usepackage{graphicx}
\usepackage{makecell}
\usepackage{mathrsfs}
\usepackage{newtxmath, newtxtext}
\usepackage{slashed}
\usepackage{subfigure}

 \usepackage[pagebackref=false, colorlinks=true]{hyperref}
\definecolor{reddish}{rgb}{0.7,0.2,0.0}  
\definecolor{blueish}{rgb}{0.1,0.1,1}
\hypersetup{
linkcolor=reddish,         
citecolor=blueish,           
filecolor=blue,              
urlcolor=magenta}       

\begin{document}

\title[Stationary Spacetimes]{Doubly Separable Spacetimes and Symmetry Constraints on their Self-Gravitating Matter Content}

\author{
Prashant Kocherlakota$^{1,2}$ and
Ramesh Narayan$^{2,1}$
}

\affiliation{
$^1$Black Hole Initiative at Harvard University, 20 Garden St., Cambridge, MA 02138, USA\\
$^2$Center for Astrophysics, Harvard \& Smithsonian, 60 Garden St., Cambridge, MA 02138, USA
}

\begin{abstract}
A popular approach to constructing exact stationary and axisymmetric \textit{nonvacuum} solutions in general relativity has been to use solution-generating techniques. Here we revisit a recent variant of the Newman-Janis-Azreg-A{\"i}nou algorithm—restricted to asymptotically-flat spacetimes—and demonstrate that this method exclusively generates Konoplya-Stuchl{\'i}k-Zhidenko spacetimes. Therefore, the equations for geodesic motion and scalar-wave propagation are both separable. We call these ``doubly separable'' spacetimes. Of these, we identify a ``degenerate'' subclass that might admit a separable Dirac equation by explicitly obtaining the Killing-Yano tensor. While the degenerate subclass is Petrov Type D, the general doubly separable spacetimes are of Type I. The high degree of symmetry in these spacetimes suggests that the self-gravitating matter must also be in specialized field configurations. For this reason, we investigate whether these spacetimes can even be sourced by arbitrary \textit{types} of matter. We show that doubly separable spacetimes cannot be sourced by massless real scalar fields or perfect fluids, and that electromagnetic fields lead only to the Kerr-Newman family. Notably, this rules out the correct spinning counterpart of the Janis-Newman-Winicour naked singularity spacetime, which contains a scalar field, as a member of this metric class. While the algorithm generates spacetimes with rich symmetry structures, valuable for studying phenomena like black hole shadows and quasinormal modes, our results highlight the need for caution when using it to construct physically consistent solutions with prespecified matter content.
\end{abstract}

\maketitle



Experimental tests of gravity conducted across a broad range of gravitational potentials and spacetime curvatures \cite{Will2014, Ferreira2019, Abbott+2016, Abbott+2017, Abbott+2021, Psaltis+2020, Kocherlakota+2021, EHTC+2022f} have firmly established general relativity (GR) as the most successful classical theory of gravity to date. As such, gaining a deeper understanding of the space of its \textit{nonvacuum} solutions and its underlying structure is both physically and theoretically valuable, e.g., for spacetime-singularity resolution \cite{Hayward2006}, addressing the cosmic-censorship conjecture \cite{Gundlach+2007}, finding the vacuum exterior spacetimes of stars with finite angular momentum \cite{Komatsu+1989}, constructing ultracompact stars that are confined states of scalar or vector fields \cite{Herdeiro+2021, Jaramillo+2023, Herdeiro+2024}, or an accurate description for black holes (BHs) embedded in massive halos \cite{Pezzella+2025, Fernandes+2025}.  

The Einstein equations describe the coupling between gravity and matter in GR, and constitute the equations of motion for the spacetime metric tensor. They form a system of nonlinear partial differential equations. To find self-consistent solutions for both the spacetime metric and its matter content, these equations must be solved simultaneously with the equations of motion for the matter fields (e.g., the Maxwell equations for an electromagnetic field).

A variety of exact solutions to the Einstein field equations have been derived for idealized physical scenarios in closed analytic form (see, e.g., Ref. \cite{Stephani+2003}). The Schwarzschild solution describes the spacetime geometry exterior to a spherically symmetric matter distribution, such as an isolated star or a non-spinning vacuum BH. The dynamical process of Schwarzschild BH formation from the spherically symmetric gravitational collapse of a homogeneous, pressureless dust cloud is captured by the Oppenheimer–Snyder solution \cite{Oppenheimer+1939}. The exterior geometry of stationary, spinning, and vacuum BHs, which are expected to form in realistic astrophysical settings, is described by the Kerr solution \cite{Kerr1963}. On cosmological scales, the large-scale structure of the universe is well modeled by the Friedmann–Lema{\^i}tre–Robertson–Walker solution \cite{Friedman1999, Lemaitre1997, Robertson1935, Walker1937}, which represents a homogeneous and isotropic universe filled with pressureless matter idealizing a collection of noninteracting galaxies, and can also contain radiation, dark matter and dark energy, in generic mixtures.

Modeling complex gravitational systems often requires going beyond the simplifying assumptions of spherical symmetry, stationarity, homogeneity, or isotropy. In such cases, analytical approaches tend to fall short and numerical methods become indispensable (see, e.g., \cite{Cardoso+2014, Lehner+2014, Dias+2016, Fernandes+2023}). For example, understanding the spacetime surrounding an isolated rotating star \cite{Stergioulas2003, Paschalidis+2016} or the dynamical formation of a Kerr BH through gravitational collapse \cite{Shibata+2002, Baiotti+2004, Giacomazzo+2011} generally relies on numerical simulations. A particularly striking example is the numerical modeling of late-stage gravitational wave emission from merging BH binaries, which has been successfully achieved within the framework of numerical relativity \cite{Pretorius2005, Campanelli+2005, Baker+2005}. 

Despite the availability of some approximate analytical techniques, there remains no general prescription for identifying a class of exact solutions corresponding to a given physical scenario (e.g., neutron star exteriors). This limitation arises largely from the intrinsic complexity of the Einstein field equations in the absence of strong symmetry assumptions. Even when a suitable metric ansatz is used (e.g., the Weyl-Lewis-Papapetrou metric \cite{Weyl2012, Lewis1932, Papapetrou1948} for vacuum stationary and axisymmetric spacetimes), solving the equations analytically remains a formidable challenge. 

To address this, and as an efficient way to obtain a more complete picture of the phase space, various analytic solution-generating techniques have been developed, primarily focused on constructing stationary and axisymmetric (spinning) spacetimes from simpler static and spherically-symmetric (nonspinning) ``seed'' solutions. These produce new solutions from known seed solutions, without having to solve the field equations afresh. 

In the case of stationary and axisymmetric electrovacuum spacetimes, the Einstein–Maxwell equations were recast as a pair of coupled complex scalar equations for the Ernst potential (a complex combination of metric functions) and a complexified electromagnetic potential \cite{Ernst1968, Ernst1968a}. These Ernst equations exhibit an integrable structure, with symmetries such as the Ehlers transformation \cite{Ehlers1993} and the Harrison transformation \cite{Harrison1968} that can be exploited to derive new solutions from known ones (reviewed recently in Ref. \cite{Vigano2022}). It is possible to cleanly find the Kerr solution from the Schwarzschild solution \cite{Ernst1968} as well as the Kerr-Newman (KN) solution from the Reissner-Nordstr{\"o}m (RN) solution \cite{Ernst1968a} in this way. The latter pair of solutions describes the spacetimes of charged black holes in GR with and without spin, respectively. The Geroch transformation \cite{Geroch1971} extends the Ehlers transformation by introducing the infinite-dimensional Geroch symmetry group, enabling a more general class of solution-generating transformations.

Almost immediately after the discovery of the Kerr solution, Newman and Janis (NJ) noticed that a series of ad hoc steps--involving a complex coordinate transformation%
\footnote{This transformation only involves the ingoing null coordinate $u$ and the radial coordinate $r$, and has remained a constant across all subsequent iterations of the NJ trick: $(u, r) \rightarrow (u^\prime, r^\prime) = (u - \mathit{i}a\cos{\vartheta}, r+\mathit{i}a\cos{\vartheta})$.
\label{fn:Complex_Coord_Transf}} %
and a decomplexification of the metric functions--could be used to transform the Schwarzschild metric into the Kerr metric \cite{Newman+1965}. Shortly thereafter, the KN metric was 
discovered by applying the NJ ``trick'' on the RN metric as the seed metric \cite{Newman+1965a}. Notably, the solution for the associated electromagnetic field could not be obtained as a product of this transformation but instead had to be solved for afterward via integration (cf., however, Ref. \cite{Erbin2015}). Nevertheless, knowledge of the metric tensor and thus the energy-momentum-stress tensor via the Einstein equations greatly simplifies this process. 

The above discussion illustrates how, in contrast to the Ernst formalism, the NJ trick is not a complete solution-generating technique. Instead, it is a \textit{metric ansatz}-generating technique that has the potential to significantly reduce the complexity of obtaining new solutions. 

Although it is possible to generate metric ansatzes for arbitrary matter fields (i.e. not just electromagnetic fields) with the NJ trick, there is no guarantee that they can be used to find a legitimate solution. Yazadjiev has shown that the spinning Kerr-Sen \cite{Sen1992} metric can be generated via the NJ trick with the nonspinning Gibbons-Maeda-Garfinkle-Horowitz-Strominger (GMGHS; \cite{Gibbons+1988, Garfinkle+1991}) metric as the seed \cite{Yazadjiev2000}. These are BH solutions to an Einstein-Maxwell-dilaton-axion theory that arises in the low-energy effective limit of the heterotic string. A compilation of other successes can be found in the modern review by Erbin \cite{Erbin2017}. The case of a minimally-coupled, massless real scalar field offers an important counterpoint. The unique static and spherically-symmetric solution there is the Janis-Newman-Winicour (JNW) naked singularity \cite{Janis+1968}. In the (discontinuous) limit in which the scalar field vanishes identically, the Schwarzschild metric is obtained. Attempts to generate a legitimate spinning generalization of the JNW metric via the NJ trick \cite{Agnese+1985} have been shown to fail \cite{Pigorov2013}. Other such failures are discussed in Ref. \cite{Azreg-Ainou2014c}. 

Of the two ambiguities of the NJ trick, the decomplexification step is perhaps more severe. It has been argued that the specific choice for the complex coordinate transformation (Footnote \ref{fn:Complex_Coord_Transf}) generalizes the seed metric by adding spin whereas other choices can generate other ``charges,'' e.g., a gravitomagnetic charge \cite{Talbot1969}. Unlike algorithmic procedures that follow a well-defined sequence of operations leading to a deterministic outcome, the decomplexification in the NJ trick often requires heuristic choices, introducing a significant degree of trial and error. Drake and Szekeres (DS) proposed a modified version of the NJ trick, replacing the decomplexification step with a requirement that the spinning metric be expressible in Boyer–Lindquist (BL; \cite{Boyer+1967}) coordinates, i.e., with only one off-diagonal component, $\mathscr{g}_{t\varphi}$ (see eq. 22 of Ref. \cite{Drake+2000}). This requirement that the generated spacetime also be \textit{circular}%
\footnote{A spacetime is called circular if the two commuting Killing vector fields $\xi^\mu = (\partial_t)^\mu$ and $\psi^\mu = (\partial_\varphi)^\mu$ are hypersurface orthogonal to a family of two-dimensional spacelike surfaces, i.e., $\xi^\alpha\mathcal{R}_\alpha^{[\beta}\xi^\gamma\psi^{\delta]} = \psi^\alpha\mathcal{R}_\alpha^{[\beta}\xi^\gamma\psi^{\delta]} = \mathbf{0}$, and $\xi_{[\alpha}\psi_\beta\nabla_\gamma\xi_{\delta]}$ and $\xi_{[\alpha}\psi_\beta\nabla_\gamma\psi_{\delta]}$ vanish at some point (see, e.g., eqs. 1-4 of Ref. \cite{Delaporte+2022}). Here $\mathcal{R}_{\alpha\beta}$ is the Ricci tensor and the square brackets indicate antisymmetrization over the enclosed indices. Thus, the metric tensor has block-diagonal structure with the $t\varphi-$sector and the $r\vartheta-$sector forming independent blocks (Theorem 7.1.1 of Ref. \cite{Wald1984}). Since a rotation in the meridional plane can always be used to set the off-diagonal metric component in the $r\vartheta-$sector to zero, circular spacetimes can always be expressed in BL coordinates (eq. 7.1.9 of Ref. \cite{Wald1984}).} %
is constraining enough to fix its form but not its relation to the initial seed metric.

Azreg-A{\"i}nou further refined the DS approach and additionally required that the generated metric reduce to the seed metric in the zero-spin limit \cite{Azreg-Ainou2014b, Azreg-Ainou2014c}. The former is then fixed by the latter, upto a conformal factor. Demanding additionally that the rest-frame of the self-gravitating matter in the generated metric coincide with the Carter tetrad (see eq. 2.10a, b of Ref. \cite{Znajek1977}) in that metric yields two partial differential equations (PDEs) for the conformal factor, which can then be solved, in principle, to completely fix the metric. For the class of ``quasi-degenerate'' seeds in particular (eq. \ref{eq:Quasi-Degenerate-Seeds-g1}), an exact solution to these PDEs was explicitly obtained (see Sec. \ref{sec:SecIC-AA-Metric}). We will refer to this series of unambiguous steps to generate a spinning metric ansatz that is uniquely associated with a nonspinning metric as the Azreg-A{\"i}nou (AA) \textit{algorithm}, and the derived metric as the AA metric. The AA algorithm has played a central role in enabling the exploration of a number of phenomenologically motivated BH models \cite{Azreg-Ainou2014c, Toshmatov+2014, Abdujabbarov+2016}. We will discuss below in Sec. \ref{sec:SecIIF-Previous-Results} why a recent attempt \cite{Solanki+2022} to generate the spinning JNW solution, now using the AA algorithm, also fails.

We have recently produced a further variation of the AA algorithm in Ref. \cite{Kocherlakota+2024}. We use the seed metric expressed in a specific set of coordinates ($\mathscr{g}_{tt}\mathscr{g}_{rr} = -1$; $\mathscr{g}_{\vartheta\vartheta}$ is the other free function) and generate a spinning metric ansatz via the AA algorithm that is completely specified, i.e., there is no ambiguity, not even a conformal factor, that one has to worry about fixing. Although this variant has been used in previous work \cite{Azreg-Ainou2016, Chen2022}, the interpretation of its matter content in Ref. \cite{Kocherlakota+2024} is new. The matter rest-frame is no longer required to be given by the Carter tetrad, as in Ref. \cite{Azreg-Ainou2014c}, but is obtained instead by solving a single \textit{quadratic} equation for the unique (timelike) angular velocity, $\Omega(r,\vartheta)$, around the spin axis of the self-gravitating matter as a function of position $(r,\vartheta)$. This is reviewed in Sec. \ref{sec:SecIIA-Matter-Rest-Frame-Review}. To better distinguish our variant from the AA algorithm, we will refer to it as the Azreg-A{\"i}nou-Chen-Kocherlakota-Narayan (ACKN) algorithm. We call the corresponding generated metric as the ACKN metric.%
\footnote{
The difference between the two algorithms can be understood as follows. Instead of demanding that the matter rest-frame be given by the Carter tetrad, our variation automatically makes an attractive choice for the conformal factor that (a) completely fixes the generated metric in terms of the seed metric without having to solve any PDEs, (b) ensures that the generated metric is asymptotically-flat if the seed metric is asymptotically-flat to begin with, and (c) ensures that the generated metric reduces exactly, instead of merely conformally, to the seed metric in the limit of zero spin.} %

While several authors have offered partial explanations for the success of the original NJ trick (see Ref. \cite{Rajan2017} and references therein), our objective here is distinct. First, we aim to characterize the class of spacetimes produced by the ACKN algorithm in terms of their symmetries and Petrov type. Spacetimes lacking these symmetries lie outside the scope of this algorithm. Second, we seek to determine the conditions under which the ACKN algorithm yields genuine solutions to the field equations, rather than merely generating metric ansatzes. This analysis will help clarify the mechanisms underlying the success of the algorithm and inform both its practical implementation and potential generalizations.

The paper is broadly organized into two parts. In Sec. \ref{sec:SecI-Separable-Spacetimes}, we will compare the ACKN metric family to three other well known metric families that arise in special contexts. 
(1) The aforementioned AA metric, which, in general, admits a separable null geodesic equation \cite{Tsukamoto+2014, Tsukamoto2018, Shaikh2019} but not necessarily a separable timelike geodesic equation or a separable scalar-wave equation. (2) The Johannsen metric family \cite{Johannsen2013}, which describes a large class of asymptotically-flat, stationary, axisymmetric and circular spacetimes that admit a separable geodesic equation for both null and timelike geodesics. (3) The Konoplya-Stuchl{\'i}k-Zhidenko (KSZ) family \cite{Konoplya+2018}, which is a subset of the Johannsen metric, that additionally supports a separable scalar-wave equation. We will show that the ACKN metric is identical to the KSZ metric, and is thus ``doubly separable,'' i.e., both the geodesic equation and the scalar-wave equation are separable. We note that this also follows from previous work by Chen and Chen \cite{Chen+2019}, who showed the DS metric to be geodesic and scalar-wave separable under certain conditions; It turns out that these conditions are ensured by the steps that reduce the DS metric to the ACKN metric. Therefore, in principle, other general results obtained for the DS metric could be used to infer various interesting properties of the ACKN metric (see Sec. \ref{sec:SecIIF-Previous-Results}).

The Killing tensor associated with the hidden symmetries allowing for the separability of the geodesic and the scalar-wave equation is discussed in Sec. \ref{sec:SecIG-Killing}. We also show that a subclass of the ACKN spacetimes (``degenerate'' ACKN spacetimes; eq. \ref{eq:Degenerate-Seed}) admit a Killing-Yano tensor, which typically implies the separability of the Dirac equation. We present the Petrov classification of the ACKN metrics in Sec. \ref{sec:SecIH-Petrov}. While generic ACKN spacetimes are of Petrov Type I, the degenerate subclass is algebraically special and is of Type D. 

In Sec. \ref{sec:SecII-Matter-in-Doubly-Separable-Metrics}, our focus will be to understand the specialized configurations of self-gravitating matter that give rise to the ACKN spacetimes with such rich symmetries. In addition to the matter rotation profiles in these spacetimes (see also Ref. \cite{Kocherlakota+2024}), such configurations also depend on the \textit{nature} of the matter fields involved. We will show that perfect fluids or massless real scalar fields cannot source these doubly separable spacetimes. The only such spacetime that can be supported by an electromagnetic field is the Kerr-Newman solution. 

Therefore, the recently obtained spinning generalization of the JNW spacetime, \textit{viz.}, the Mirza–Kangazi–Sadeghi (MKS) solution \cite{Mirza+2023}, lies outside the class of ACKN metrics. The MKS solution provides an explicit, closed-form expression for both the metric and the accompanying self-gravitating scalar field, and verifies that the configuration satisfies the full Einstein–Klein–Gordon system. This represents a significant advance, arriving more than five decades after the original static solution. Ref. \cite{Mirza+2023} also analyzes the MKS solution within the Ernst solution-generating formalism, explicitly deriving the corresponding Ernst potential. It is further argued that the solution cannot be obtained via a Newman–Janis-type algorithm, consistent with our findings.

This result on electromagnetic fields has been established for the DS metric by looking for those solutions that admit a trace-free Ricci tensor \cite{Drake+2000}, and is now understood to be inherited by the ACKN metric. 

We emphasize that the methods described below can be readily generalized to other matter fields, enabling an investigation into whether such symmetries are compatible with arbitrary types of self-gravitating matter models. This, in turn, may provide valuable insights toward constructing an exact interior solution for the Kerr metric as well as for extending the compactness bounds in GR \cite{Alho+2022}. 

We adopt Geometrized units in which $G=c=1$. For each stationary and axisymmetric metric, $a$ is the spin parameter, which measures the rotation of the spacetime. When $a=0$, each reduces to a static and spherically-symmetric spacetime. While our results are not limited to black hole spacetimes, their application to such cases pertains only to the exterior region. Extending the analysis to include the interior cosmological region is straightforward.

\section{Spacetimes with Hidden Symmetries} 
\label{sec:SecI-Separable-Spacetimes}

We start by providing a brief overview of the different classes of spacetime metrics that we are interested in here, following which we describe the relationship between them in Sec. \ref{sec:SecIF-Metric-Relations}. 

As outlined above, the spinning Azreg-A{\"i}nou–Chen–Kocherlakota–Narayan (ACKN) and the spinning Azreg-A{\"i}nou (AA) metrics are obtained from general nonspinning seed metrics via the ACKN or the AA algorithms. Before discussing each of these metric families, we begin by reviewing equivalent formulations of the most general seed metric, along with certain special subclasses that will recur in various contexts in this work.


\subsection{Nonspinning Metrics}
\label{sec:IIA-Nonspinning-Metrics}

In general spherical-polar coordinates, $x^\mu = (t, r, \vartheta, \varphi)$, an arbitrary static and spherically-symmetric seed metric takes the form, 
\begin{equation} \label{eq:Seed-Metric-OD}
\mathrm{d}s^2 = -f(r)\mathrm{d}t^2 + \frac{g(r)}{f(r)}\mathrm{d}r^2 + R^2(r)\mathrm{d}\Omega_2^2\,,
\end{equation}
where $\mathrm{d}\Omega_2^2 = \mathrm{d}\vartheta^2 + \sin^2{\vartheta}\mathrm{d}\varphi^2$. In the above, the seed metric is described using three free functions. However, a redefinition of the radial coordinate can always be used to eliminate one of the functions $g(r)$ or $R(r)$. The function $f(r)$ cannot be eliminated by coordinate transformations without introducing off-diagonal metric components such as $\mathscr{g}_{tr}$ (see eq. 14.19 of Ref. \cite{Plebanski+2012}). For this reason, we refer to eq. \ref{eq:Seed-Metric-OD} as an ``over-determined'' form of a generic seed metric.

The metric function $g(r)$ can be eliminated in eq. \ref{eq:Seed-Metric-OD} via a transformation $r \mapsto \rho$, where $\rho$ is a solution of the ordinary differential equation $\mathrm{d}\rho = \sqrt{g(r)}\mathrm{d}r$. The seed metric \eqref{eq:Seed-Metric-OD} then becomes
\begin{equation}
\label{eq:Seed-Metric-g1-v1}
\mathrm{d}s^2 = -f(\rho)\mathrm{d}t^2 + \frac{\mathrm{d}\rho^2}{f(\rho)} + R^2(\rho)\mathrm{d}\Omega_2^2\,.   \end{equation} 

Alternatively, the metric function $R(r)$ can be eliminated in eq. \ref{eq:Seed-Metric-OD} by choosing to rewrite $g(r)$ in terms of a new function $\tilde{g} = g/(\partial_rR)^2$. Then
\begin{align}
\label{eq:Seed-Metric-Rr-v1}
\mathrm{d}s^2 = -f(R)\mathrm{d}t^2 + \frac{\tilde{g}(R)}{f(R)}\mathrm{d}R^2 + R^2\mathrm{d}\Omega_2^2\,.
\end{align}

If we simply relabel the radial coordinate in both eqs. \ref{eq:Seed-Metric-g1-v1} and \ref{eq:Seed-Metric-Rr-v1} by $r$, and also relabel $\tilde{g}$ by $g$ in eq. \ref{eq:Seed-Metric-Rr-v1}, then the two uniquely-defined forms of the seed metric are
\begin{align} 
\label{eq:Seed-Metric-g1}
\mathrm{g=1\ Form}:&\ \mathrm{d}s^2 = -f(r)\mathrm{d}t^2 + \frac{\mathrm{d}r^2}{f(r)} + R^2(r)\mathrm{d}\Omega_2^2\,, \\
\label{eq:Seed-Metric-Rr}
\mathrm{R=r\ Form}:&\ \mathrm{d}s^2 = -f(r)\mathrm{d}t^2 + \frac{g(r)}{f(r)}\mathrm{d}r^2 + r^2\mathrm{d}\Omega_2^2\,.
\end{align} 
We refer to the choice of the coordinates that yield the $g=1$ form \eqref{eq:Seed-Metric-g1} as the $g=1$ coordinates and those that produce the $R=r$ form as areal/curvature coordinates.

In Sec. \ref{sec:SecII-Matter-in-Doubly-Separable-Metrics}, we will almost exclusively work in areal/curvature coordinates for the seed spherically-symmetric metric and in associated coordinates for the generated axisymmetric metric. The only exception will be discussion surrounding eq. \ref{eq:pth-Equal-pr} in Sec. \ref{sec:SecIID-Meridional}, for which we found it best to adopt $g=1$ coordinates.

Specifying \textit{two} of the metric functions simultaneously corresponds to picking out a specific class of static and spherically-symmetric spacetimes. Two such classes occupy an interesting role in the space of spacetime geometries: the degenerate (see, e.g., Ref. \cite{Jacobson2007} for further discussion) and the quasi-degenerate metrics.

Spacetimes that are described by the metric \eqref{eq:Seed-Metric-OD} with both $R(r) = r$ \textit{and} $g(r) = 1$ are referred to as ``degenerate'' seed metrics. This is because they depend on a single free function,
\begin{align} \label{eq:Degenerate-Seed}
\mathrm{Degenerate\ Seeds}:\ \mathrm{d}s^2 = -f(r)\mathrm{d}t^2 + \frac{\mathrm{d}r^2}{f(r)} + r^2\mathrm{d}\Omega_2^2\,.
\end{align}
When referring to degenerate spacetimes below, we will also mean the ACKN spacetimes that are generated from degenerate nonspinning seeds. The Kerr-Newman family \cite{Newman+1965a} belongs to the class of degenerate spacetimes.

We now introduce the class of ``quasi-degenerate'' seed metrics—a closely related one-parameter ($r_0$) generalization of the degenerate  metrics. For these spacetimes, the metric is of the form
\begin{align}
 \label{eq:Quasi-Degenerate-Seeds-g1}
&\ \mathrm{Quasi\!-\!Degenerate\ Seeds\ (g=1\ Form)}: \nonumber \\
&\ \quad \mathrm{d}s^2 = -f(r)\mathrm{d}t^2 + \frac{\mathrm{d}r^2}{f(r)} + \left(r^2 - r_0^2\right)\mathrm{d}\Omega_2^2\,, \\
 \label{eq:Quasi-Degenerate-Seeds-Rr}
&\ \mathrm{Quasi\!-\!Degenerate\ Seeds\ (R=r\ Form)}: \nonumber \\
&\ \quad \mathrm{d}s^2 = -f(r)\mathrm{d}t^2 + \left(\frac{r^2}{r^2+r_0^2}\right)\frac{\mathrm{d}r^2}{f(r)} + r^2\mathrm{d}\Omega_2^2\,.
\end{align}
Again, when referring to quasi-degenerate spacetimes below, we will also mean the ACKN spacetimes that are generated from the quasi-degenerate seeds. The Kerr-Sen family \cite{Sen1992} belongs to the class of (non-degenerate) quasi-degenerate spacetimes. For the Kerr-Sen family, we have $r_0 = Q^2/(2M)$, where $M$ and $Q$ denote the mass and charge of the spacetime (see the expression for the metric function $B^2$ for the metric labeled ``EMd'' in Table I of Ref. \cite{Kocherlakota+2020}).

\subsection{The Azreg-A{\"i}nou-Chen-Kocherlakota-Narayan Metric} 
\label{sec:SecIB-ACKN-Metric}

The spinning ACKN metric generated from the seed metric \eqref{eq:Seed-Metric-OD} is given, in BL coordinates, $x^\mu = (t,r,\vartheta,\varphi)$, as (see, e.g., eq. 2.4 of Ref. \cite{Kocherlakota+2024})
\begin{align} \label{eq:ACKN-Metric}
\mathrm{d}s^2 =&\ -\left(1-\frac{2F}{\Sigma}\right)\mathrm{d}t^2 -2\frac{2F}{\Sigma}a\sin^2{\vartheta}~\mathrm{d}t\mathrm{d}\varphi + \frac{\Pi}{\Sigma}\sin^2{\vartheta}~\mathrm{d}\varphi^2 \nonumber \\
&\ + \frac{\Sigma}{\Delta}g~\mathrm{d}r^2 + \Sigma~\mathrm{d}\vartheta^2\,,
\end{align}
where the generated metric functions, $F, \Delta, \Sigma$ and $\Pi$, are related to the seed metric functions, $f$ and $R$, via
\begin{align}
2F(r) &=\ (1-f)R^2\,; \\ 
\Delta(r) &=\ fR^2+a^2\,, \nonumber\\
\Sigma(r, \vartheta) &=\ R^2+a^2\cos^2{\vartheta}\,, \nonumber\\
\Pi(r, \vartheta) &=\ (R^2+a)^2 - \Delta a^2\sin^2{\vartheta}\,. \nonumber
\end{align}
The ACKN metric \eqref{eq:ACKN-Metric} admits a separable geodesic equation for both timelike and null geodesics. As we will see below, it also admits a separable scalar-wave equation.


\subsection{The Azreg-A{\"i}nou Metric} 
\label{sec:SecIC-AA-Metric}

We mentioned above that the ACKN metric differs nontrivially from the original AA metric \cite{Azreg-Ainou2014a, Azreg-Ainou2014c}. We now briefly describe the AA metric for easy comparison.

For the general static and spherically-symmetric seed metric in equation \ref{eq:Seed-Metric-OD}, the associated $a\neq 0$ AA metric is given, in BL coordinates, as (see, e.g., eq. G.1 of Ref. \cite{Kocherlakota+2024})
\begin{align} \label{eq:AA-Metric}
\mathrm{d}s^2
=&\ 
\frac{\hat{X}}{\hat{\Sigma}}\left[-\left(1-\frac{2 \hat{F}}{\hat{\Sigma}}\right)\mathrm{d}t^2 
-2\frac{2 \hat{F}}{\hat{\Sigma}}a\sin^2{\vartheta}~\mathrm{d}t\mathrm{d}\varphi\right. \\ 
&\ 
\left. +\frac{\hat{\Pi}}{\hat{\Sigma}}\sin^2{\vartheta}~\mathrm{d}\varphi^2 + \frac{\hat{\Sigma}}{\hat{\Delta}}\mathrm{d}r^2 + \hat{\Sigma}~\mathrm{d}\vartheta^2\right]\,, \nonumber
\end{align}
where the metric functions, $\hat{F}, \hat{\Delta}, \hat{\Sigma}$ and $\hat{\Pi}$, are related to the seed metric functions, $f, g$ and $R$, via
\begin{align}
2\hat{F}(r) &=\ (1-f/\sqrt{g})R^2/\sqrt{g}\,; \\ 
\hat{\Delta}(r) &=\ (f/g)R^2+a^2\,, \nonumber\\
\hat{\Sigma}(r, \vartheta) &=\ R^2/\sqrt{g}+a^2\cos^2{\vartheta}\,, \nonumber\\
\hat{\Pi}(r, \vartheta) &=\ (R^2\sqrt{g}+a)^2 - \hat{\Delta} a^2\sin^2{\vartheta}\,. \nonumber
\end{align}
A comparison of eq. \ref{eq:ACKN-Metric} and eq. \ref{eq:AA-Metric} reveals two differences: The presence of $g(r)$ in the former, and the factor $\hat{X}/\hat{\Sigma}$ in the latter. The metric function $\hat{X}$ (the conformal factor) in the AA metric is fixed, in principle, by the requirement that the rest-frame of the matter in these spacetimes coincide with the Carter tetrad (see Sec. 3 of Ref. \cite{Azreg-Ainou2014a}). The Carter tetrad is given by the tetrad in eq. \ref{eq:Rest-Frame} below with (see Ref. \cite{Znajek1977} for the Kerr Carter tetrad) a $\vartheta$-independent angular velocity (i.e., rigid rotation at each radius),
\begin{equation} \label{eq:Carter-Tetrad}
\Omega = \frac{a}{(R^2/\sqrt{g})+a^2}\,.
\end{equation}
More specifically, requiring that the $r\vartheta-$ and the $t\varphi-$components respectively of the Einstein tensor for the AA metric vanish identically in the Carter tetrad leads to two partial differential equations (PDEs) for $\hat{X}$ (see eqs. 4, 7 of Ref. \cite{Azreg-Ainou2014a}),
\begin{align} \label{eq:AA_PDEs}
0 =&\ \sqrt{\hat{X}}\partial_r\partial_y\left(1/\sqrt{\hat{X}}\right) - \sqrt{\hat{\Sigma}}\partial_r\partial_y\left(1/\sqrt{\hat{\Sigma}}\right)\,, \\
0 =&\ \hat{X}\left[-\hat{\Sigma}\partial_r^2\hat{A} + \left(\partial_r \hat{A}\right)^2 + 2\left(\hat{A} - a^2y^2\right)\right] \nonumber  \\ 
&\ + \Sigma\left[2y\partial_y\hat{X} - \partial_r \hat{A} \cdot \partial_r \hat{X}\right]\,. \nonumber 
\end{align}
In the above, we have introduced $y=\cos{\vartheta}$ and $\hat{A} = R^2/\sqrt{g}$ for brevity. While $\hat{X}=\hat{\Sigma}$ is clearly a solution of the first equation, there is no guarantee that it also solves the second. Setting $\hat{X}=\hat{\Sigma} = \hat{A} + a^2y^2$ in the second equation 
reduces it to an ordinary differential equation for $\hat{A}$ 
\begin{equation} \label{eq:X-Equal-Sigma-Eq}
\mathrm{For}\ \hat{X}=\hat{\Sigma}:\ \partial_r^2 \hat{A} = 2\,,
\end{equation}
i.e., (see also eq. 8 of Ref. \cite{Azreg-Ainou2014a})
\begin{equation}
\hat{A}(r) = r^2 + p_1 r + p_0 = (r+r_1)^2 - r_0^2\,.
\end{equation}
The first form of $\hat{A}$ above is evidently the solution to eq. \ref{eq:X-Equal-Sigma-Eq}, with $p_1$ and $p_0$ some integration constants. We can rewrite these constants in terms of two other constants $r_1$ and $r_0$ as $p_1 = 2r_1$ and $p_0 = r_1^2 - r_0^2$ to obtain the second version of $\hat{A}$ above. Adopting coordinates in which $g(r) = 1$, and shifting the radial coordinate by the constant $r_1$, we identify this to be the class of quasi-degenerate seeds \eqref{eq:Quasi-Degenerate-Seeds-g1}. 

The result above implies that spinning AA spacetimes not generated from quasi-degenerate seeds cannot satisfy the condition $\hat{X} = \hat{\Sigma}$. Consequently, quasi-degenerate seeds (which include as a sub-class the degenerate seeds) define the unique class of spacetimes shared by both the ACKN and AA constructions. There is, however, one caveat. As we will discuss in Sec. \ref{sec:SecIIB-Rigid-Rotation}, which focuses on ACKN spacetimes admitting rigidly rotating matter, there exists an unusual class of seed spacetimes characterized by $f = 1$ that give rise to axisymmetric geometries without frame-dragging (i.e., with $\mathscr{g}_{t\varphi} = 0$). This constitutes the second class of spacetimes common to both the ACKN and AA families.

We note also that for solutions to the PDEs \eqref{eq:AA_PDEs} $\hat{X}\neq\hat{\Sigma}$, the zero-spin limit of the AA metric \eqref{eq:AA-Metric} is, in general, only conformally-related to the spherically-symmetric seed metric \eqref{eq:Seed-Metric-OD} that was used as an input (see Sec. 4 of Ref. \cite{Azreg-Ainou2014a}). This is also true for the quasi-degenerate metrics except in $g=1$ coordinates.

Finally, we reiterate that the AA metric in general admits only a separable null geodesic equation \cite{Tsukamoto+2014, Tsukamoto2018, Shaikh2019}. When the solution to eq. \ref{eq:AA_PDEs}, $\hat{X}(r, \vartheta)$, is of the form $\hat{X}(r, \vartheta) = \hat{X}_r(r) + \hat{X}_\vartheta(\vartheta)$, then the timelike geodesic equation is also separable \cite{Chen+2019, Kocherlakota+2023}. 

\subsection{The Johannsen Metric} 
\label{sec:SecID-Johannsen-Metric}

The Johannsen metric was constructed in Ref. \cite{Johannsen2013} by modifying the Kerr metric while preserving the separability of the geodesic equation. To the best of our knowledge, there is currently no proof that the Johannsen metric represents the most general class of stationary, axisymmetric, and circular spacetimes that both admit a separable geodesic equation and include the Kerr metric as a special case.

The Johannsen metric allows for four free functions of the radial coordinate, $\bar{f}(\bar{r}), A_1(\bar{r}), A_2(\bar{r})$ and $A_5(\bar{r})$. In BL coordinates, $x^{\bar{\mu}}=(\bar{t}, \bar{r}, \bar{\vartheta}, \bar{\varphi})$, the Johannsen metric takes the form (eq. 51 of Ref. \cite{Johannsen2013})
\begin{align}
\mathrm{d}s^2 =&\ 
-\frac{\bar{\Sigma}}{\bar{X}^2}\left(\bar{\Delta} - A_2^2a^2\sin^2{\bar{\vartheta}}\right)\mathrm{d}\bar{t}^2 \\
&\ -2\frac{\bar{\Sigma}}{\bar{X}^2}\left[(\bar{r}^2+a^2)A_1A_2 - \bar{\Delta}\right]a\sin^2{\bar{\vartheta}}~\mathrm{d}\bar{t}\mathrm{d}\bar{\varphi} 
\nonumber \\
&\ + \frac{\bar{\Sigma}}{\bar{X}^2}\bar{\Pi}\sin^2{\bar{\vartheta}}~\mathrm{d}\bar{\varphi}^2 
+ \frac{\bar{\Sigma}}{\bar{\Delta}A_5}\mathrm{d}\bar{r}^2 + \bar{\Sigma}~\mathrm{d}\bar{\vartheta}^2\,, \nonumber
\end{align}
where the functions appearing above are given in terms of the four free functions via the relations
\begin{align}
\bar{\Delta}(\bar{r}) =&\ \bar{r}^2 - 2Mr + a^2\,, \\
\bar{\Sigma}(\bar{r}, \bar{\vartheta}) =&\ \bar{f} + \bar{r}^2 + a^2\cos^2{\bar{\vartheta}}\,; \nonumber \\
\bar{\Pi}(\bar{r}, \bar{\vartheta}) =&\ (\bar{r}^2 + a^2)^2A_1^2 - \bar{\Delta} a^2 \sin^2{\bar{\vartheta}}\,, \nonumber \\
\bar{X}(\bar{r}, \bar{\vartheta}) =&\ (\bar{r}^2 + a^2)A_1 - A_2 a^2 \sin^2{\bar{\vartheta}}\,. \nonumber
\end{align}
The zero spin limit of the Johannsen metric is
\begin{equation} \label{eq:Johannsen-a0}
\mathrm{d}s^2 =
-\left(1-\frac{2M}{\bar{r}}\right)\frac{\bar{\Sigma}_0}{\bar{r}^2A_1^2}\mathrm{d}\bar{t}^2 + \left(1-\frac{2M}{\bar{r}}\right)^{-1}\frac{\bar{\Sigma}_0}{\bar{r}^2A_5}\mathrm{d}\bar{r}^2 + \bar{\Sigma}_0~\mathrm{d}\bar{\Omega}_2^2\,,
\end{equation}
where $\bar{\Sigma}_0 = \bar{f}+\bar{r}^2$ and $\mathrm{d}\bar{\Omega}_2^2 = \mathrm{d}\bar{\vartheta}^2 + \sin^2{\bar{\vartheta}}\mathrm{d}\bar{\varphi}^2$ is the standard line element on a unit $2-$sphere. The function $A_2$ does not appear in equation \ref{eq:Johannsen-a0}, so the zero-spin version of the Johannsen metric involves only three functions $\bar{f}, A_1$ or $A_5$. Further, since an arbitrary static and spherically-symmetric metric can be described by just two functions (eq. \ref{sec:IIA-Nonspinning-Metrics}, see e.g., Ch. 14 of Ref. \cite{Plebanski+2012}), it is clear that one of the functions $\bar{f}, A_1$ or $A_5$ can be eliminated by a change of the radial coordinate. For example, $A_5=\bar{\Sigma}_0^2/(r^2A_1^2)$ or $\mathscr{g}_{\bar{t}\bar{t}}\mathscr{g}_{\bar{r}\bar{r}}=-1$ is always a valid choice. 


\subsection{The Konoplya-Stuchl{\'i}k-Zhidenko Metric} 
\label{sec:SecIE-KSZ-Metric}

The Konoplya-Stuchl{\'i}k-Zhidenko (KSZ) metric \cite{Konoplya+2018} is obtained from the general stationary, axisymmetric, and circular Konoplya-Rezzolla-Zhidenko (KRZ) metric \cite{Konoplya+2016} by demanding that the spacetime admit a separable geodesic as well as scalar-wave equation. The symmetry generators underlying such separability are discussed below in Sec. \ref{sec:SecIG-Killing}. A related notion of ``ladder symmetries'' of the general KRZ spacetime has recently been explored in Ref. \cite{Sharma+2024}.

The KSZ metric allows for three free functions of the radial coordinate, $R_\Sigma, R_B$ and $R_M$. In BL coordinates, $x^{\tilde{\mu}} = (\tilde{t}, \tilde{r}, \tilde{\vartheta}, \tilde{\varphi})$, the metric takes the form (eqs. 1 and 22 of Ref. \cite{Konoplya+2018})
\begin{align}
\mathrm{d}s^2 =&\ -\left(\frac{N^2-W^2\sin^2{\tilde{\vartheta}}}{K^2}\right)\mathrm{d}\tilde{t}^2 -2\tilde{r}W\sin^2{\tilde{\vartheta}}~\mathrm{d}\tilde{t}\mathrm{d}\tilde{\varphi} \nonumber \\
&\ + \tilde{r}^2K^2\sin^2{\tilde{\vartheta}}~\mathrm{d}\tilde{\varphi}^2 + \tilde{\Sigma}\frac{B^2}{N^2}\mathrm{d}\tilde{r}^2 + \tilde{r}^2\tilde{\Sigma}~\mathrm{d}\tilde{\vartheta}^2\,,
\end{align}
where the functions appearing above are given in terms of the three free functions via the relations
\begin{align}
\tilde{\Sigma}(\tilde{r}, \tilde{\vartheta}) =&\ R_\Sigma + \frac{a^2\cos^2{\tilde{\vartheta}}}{\tilde{r}^2}\,, \\
B(\tilde{r}) =&\ R_B\,, \nonumber\\
N^2(\tilde{r}) =&\ R_\Sigma - \frac{R_M}{\tilde{r}} + \frac{a^2}{\tilde{r}^2}\,, \nonumber \\
W(\tilde{r}, \tilde{\vartheta}) =&\ \frac{R_M}{\tilde{\Sigma}}\frac{a}{\tilde{r}^2}\,, \nonumber \\
K^2(\tilde{r}, \tilde{\vartheta}) =&\ \frac{1}{\tilde{\Sigma}}\left[R^2_\Sigma + R_\Sigma\frac{a^2}{\tilde{r}^2} + \frac{N^2}{r^2}a^2\cos^2{\tilde{\vartheta}}\right] + \frac{a W}{\tilde{r}}\,. \nonumber
\end{align}
The zero spin limit of the KSZ metric is
\begin{equation}
\mathrm{d}s^2 =
-\left(1-\frac{R_M}{\tilde{r}R_\Sigma}\right)\mathrm{d}\tilde{t}^2 + \left(1-\frac{R_M}{\tilde{r}R_\Sigma}\right)^{-1}R_B^2~\mathrm{d}\tilde{r}^2 + \tilde{r}^2 R_\Sigma~\mathrm{d}\tilde{\Omega}_2^2\,, \nonumber
\end{equation}
where $\mathrm{d}\tilde{\Omega}_2^2 = \mathrm{d}\tilde{\vartheta}^2 + \sin^2{\tilde{\vartheta}}\mathrm{d}\tilde{\varphi}^2$. As above, one of the three metric functions can always be eliminated.


\subsection{Relating the ACKN, Johannsen and KSZ Metrics} 
\label{sec:SecIF-Metric-Relations}

To minimize the complexity of comparing the ACKN metric with the Johannsen and KSZ metrics, it is useful to work in BL coordinates, which admit only one nontrivial off-diagonal component ($\mathscr{g}_{t\varphi}$). We will come back to the relation between the ACKN and the AA metrics in Sec. \ref{sec:SecIIB-Rigid-Rotation} below.

Our strategy here is as follows. When comparing the ACKN metric to the Johannsen metric, we will transform the ACKN metric into the form it takes in the specific set of BL coordinates that the Johannsen metric is written in, i.e., we perform a coordinate transformation $x^\mu \mapsto x^{\bar{\mu}}$. Since both of these coordinate systems are BL systems, the Jacobian for this coordinate transformation, $\Lambda^\mu_{\ \hat{\mu}}=\partial_{\hat{\mu}}x^\mu$, must necessarily be of the form (see the discussion in Appendix \ref{AppA:BL-Systems})
\begin{align*}
\Lambda^\mu_{\ \hat{\mu}} = 
\begin{bmatrix}
\alpha & 0 & 0 & \beta \\
0 & \chi & 0 & 0 \\
0 & 0 & \xi & 0 \\
\gamma & 0 & 0 & \delta
\end{bmatrix}\,,
\end{align*}
where $\alpha, \beta, \gamma,$ and $\delta$ are constants. Furthermore, we have defined $\chi = \partial_r \bar{r}$ and $\xi = \partial_\vartheta \bar{\vartheta}$. We then compute the difference between each metric component of the ACKN and the Johannsen metric (now in the same set of coordinates). Demanding that this difference vanish yields maps between the various free functions, i.e., the Johannsen free functions are given in terms of the ACKN free functions. If this difference only vanishes for a specific combination of the Johannsen free functions, then we can conclude that the ACKN metric is a subclass of the Johannsen metric. 

Demanding that the difference between the $\bar{\vartheta}\bar{\vartheta}-$component of the Johannsen and ACKN metric vanish admits a simple solution: $\xi = 1$ or, equivalently, $\bar{\vartheta}(\vartheta) = \vartheta$, and $\bar{\Sigma} = \Sigma$, i.e.,
\begin{equation}
\label{eq:barf}
\bar{f} = R^2 - \bar{r}^2\,.    
\end{equation}
Demanding further that the difference between the three independent components of the $\bar{t}\bar{\varphi}-$sector vanish yields the unique solution $\{\alpha, \beta, \gamma, \delta\} = \{1,0,0,1\}$ and that 
\begin{equation}
\label{eq:A1A2}
A_1 = \sqrt{\frac{\bar{\Delta}}{\Delta}}\frac{R^2 + a^2}{\tilde{r}^2 + a^2}\,;\ 
A_2 = \sqrt{\frac{\bar{\Delta}}{\Delta}}\,.
\end{equation}
Finally, demanding that the difference between $\mathscr{g}_{\bar{r}\bar{r}}$ in the ACKN and Johannsen metrics vanish admits a simple solution: $\chi = 1$ or, equivalently, $\bar{r}(r) = r$, and
\begin{equation}
\label{eq:A5}
A_5 = \frac{\Delta}{\bar{\Delta}}\frac{1}{g}\,.
\end{equation}
Since we have found the Jacobian for the coordinate transformation $x^\mu \mapsto x^{\bar{\mu}}$ to be the identity matrix, we will drop the overbars from the coordinates. The three equations above can be understood as follows. Picking the metric function $R$ in the ACKN metric determines $\bar{f}$ in the Johannsen metric via eq. \ref{eq:barf}. Then picking $f$ in the ACKN metric determines $A_2$ in the Johannsen metric via the second relation in eq. \ref{eq:A1A2}. Finally, picking $g$ in the ACKN fixes $A_5$ in the Johannsen metric via eq. \ref{eq:A5}. Thus, the relation for $A_1$ in eq. \ref{eq:A1A2} is actually a constraint equation on the Johannsen metric functions of the form,
\begin{equation} \label{eq:Constraint}
\bar{f} = (r^2+a^2)\left(\frac{A_1}{A_2} - 1\right)\,.
\end{equation}
The ACKN metric is, therefore, a proper subset of the Johannsen metric, viz., the subset that satisfies eq. \ref{eq:Constraint}.

It is worth noting here that Ref. \cite{Salehi+2023} has previously demonstrated that the class of degenerate ACKN metrics \eqref{eq:Degenerate-Seed} is a subclass of the Johannsen metric. Our result is more general.

We repeat the same process for the ACKN metric and the KSZ metric but now with a coordinate transformation $x^\mu \mapsto x^{\tilde{\mu}}$. We find the Jacobian relating the two coordinate systems to again be the identity matrix and obtain the following map between the two sets of metric functions,
\begin{equation}
R_\Sigma = \frac{R^2}{r^2}\,;\ R_B = \sqrt{g}\,;\ R_M = \frac{(1-f)R^2}{r}\,.
\end{equation}
The three functions $R_\Sigma, R_B, R_M$ in the KSZ metric are thus uniquely determined by the three free functions, $f, g, R$, in the ACKN metric, and vice versa. Thus, the ACKN metric is identical to the KSZ metric. This is not surprising in light of eq. \ref{eq:Constraint}: Ref. \cite{Konoplya+2018} has established the KSZ metric to be that specific proper subset of the Johannsen metric which satisfies this equation (see eq. 38 there). 

\begin{figure}
\centering
\includegraphics[width=\linewidth]{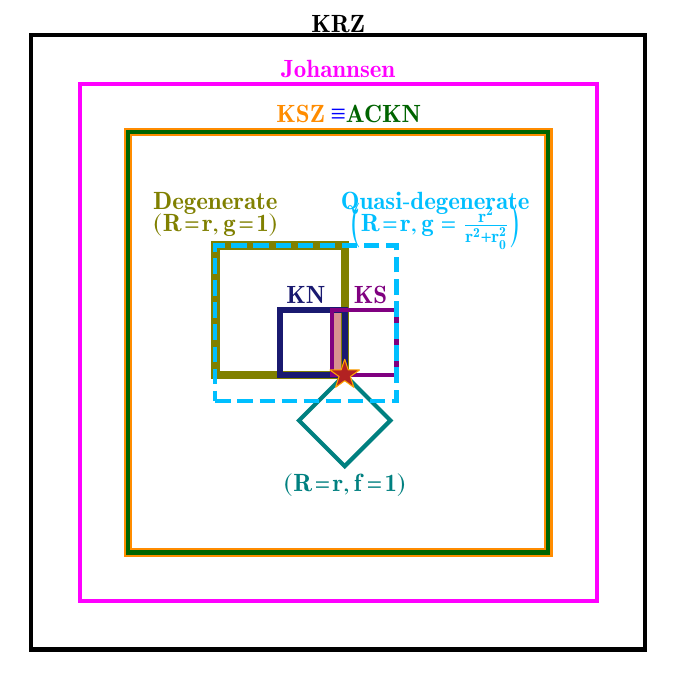}
\caption{
A Venn diagram for a pictoral representation of the relations between the various metrics discussed in this paper. Included here are the Konoplya-Rezzolla-Zhidenko (KRZ), the Johannsen, the Konoplya-Stuchl{\'i}k-Zhidenko (KSZ), and the Azreg-A{\"i}nou-Chen-Kocherlakota-Narayan (ACKN) metric families. The Azreg-A{\"i}nou (AA) metric family is not shown, as its relation to the other spacetimes is not fully understood. However, the AA metric and the ACKN metric family have the Quasi-degenerate ($R=r, g=r^2/(r^2+r_0^2)$) and a strange family of nonspinning and axisymmetric spacetimes ($R=r, f=1$) in common, as discussed in Sec. \ref{sec:SecIC-AA-Metric} and Sec. \ref{sec:SecIIB-Rigid-Rotation}. The Johannsen spacetimes admit a Killing tensor due to the separability of the geodesic equation. Only the subclass of ACKN/KSZ spacetimes also admit a separable scalar-wave equation. Of these, the degenerate ACKN/KSZ spacetimes admit a Killing-Yano tensor and, therefore, likely also a separable Dirac equation (Sec. \ref{sec:SecIG-Killing}). The ACKN/KSZ spacetimes are in general of Petrov Type I but the subclass of degenerate spacetimes are of Type D (Sec. \ref{sec:SecIH-Petrov}). Shown also are the Kerr-Newman (KN) and the Kerr-Sen (KS) \textit{solution} families. Finally, the Kerr metric is shown in red shading and the star represents the Minkwoski spacetime. The sizes of the shapes hold no meaning.}
\label{fig:Fig1-Venn}
\end{figure}

Fig. \ref{fig:Fig1-Venn} shows a pictoral representation of the relationships between the various metric families considered here and in the remainder of this paper.

\subsection{Killing and Killing-Yano Tensors in Doubly Separable Spacetimes} 
\label{sec:SecIG-Killing}


In an ACKN spacetime, the particle $4-$momentum, $\mathbf{p}=p_\mu\mathrm{d}x^\mu$, which satisfies the geodesic equation 
\begin{equation} \label{eq:Geodesic-Equation}
p_\mu \nabla^\mu p_\nu = 0\,,
\end{equation}
is additively separable, i.e., it takes the form
\begin{equation} \label{eq:Principal-Function}
\mathbf{p} = -E \mathrm{d}t \pm_r E\frac{g}{\Delta}\sqrt{\mathscr{R}}\mathrm{d}r \pm_\vartheta E\sqrt{\Theta}\mathrm{d}\vartheta + L \mathrm{d}\varphi\,,
\end{equation}
where the radial and polar potentials are given respectively as
\begin{align}
\mathscr{R}(r) =&\ \frac{\left[(R^2+a^2) - a\xi\right]^2 - \Delta\eta^2 - m^2 E^{-2}R^2\Delta}{g}\,,  \\
\Theta(\vartheta) =&\ \eta^2 - (a\sin{\vartheta} - \xi\csc{\vartheta})^2 -m^2E^{-2}a^2\cos^2{\vartheta}\,.
\end{align}
In the above, $\xi= L/E$ and $\eta = \sqrt{C}/E$ are the two impact parameters of the geodesic. 

In such spacetimes, the geodesic equation reduces to a system of first-order ordinary differential equations, and the particle coordinate $4-$velocity, $\dot{x}^\mu = \mathrm{d}x^\mu/\mathrm{d}\lambda_{\rm m}$,%
\footnote{Here $\lambda_{\rm m}$ is the Mino time \cite{Mino2003} along the geodesic.} %
\begin{equation} \label{eq:Geodesic-Tangent}
\dot{x}^\mu = [\mathscr{T}_r + a^2\cos^2{\vartheta}, \pm_r \sqrt{\mathscr{R}}, \pm_\vartheta\sqrt{\Theta}, \Phi_r + \xi\csc^2{\vartheta}]\,,
\end{equation}
where
\begin{align}
\mathscr{T}_r(r) =&\ \frac{(R^2+a^2)^2 - 2Fa\xi}{\Delta} - a^2\,,\ \Phi_{r}(r) =&\ \frac{a(2F - a\xi)}{\Delta}\,, \nonumber
\end{align}
can be directly integrated to obtain the complete particle orbit, $x^\mu(\lambda_{\rm m})$. Shadows of KSZ black holes have been studied in Ref. \cite{Konoplya+2021}. For the properties of shadows, the photon shells and the photon rings in ACKN spacetimes, see Ref. \cite{Walia+2025}. 

The complete separability of the geodesic equation \eqref{eq:Principal-Function} implies the existence of four constants of the motion. Two of these, $E = -p_\mu(\partial_t)^\mu$ and $L = +p_\mu(\partial_\varphi)^\mu$, correspond to the conserved energy and azimuthal angular momentum along the geodesic respectively, which are associated with the two Killing vectors of the spacetime ($\partial_t$ and $\partial_\varphi$). The remaining conserved quantities, $m$ and $C$, are associated with the existence of two rank-two Killing \textit{tensors}.

Any continuous symmetry of the spacetime manifold gives rise to a corresponding symmetry on the phase space of geodesics (because the latter is a subset of the cotangent bundle of the former%
\footnote{The region that satisfies the particle mass on-shell/Hamiltonian constraint.}). %
When ``lifted up'' to phase space, these symmetries lead to conserved quantities that are linear in the canonical momenta ($E, L$). However, the reverse is not necessarily true: not all symmetries of the phase space originate from transformations of the underlying spacetime. Symmetries that do correspond to spacetime isometries are called \textit{explicit} symmetries, while those that do not have a direct counterpart in the manifold or the configuration space are referred to as \textit{hidden} symmetries \cite{Frolov+2017}.

A Killing tensor, $\mathscr{K}^{\mu\nu\cdots\zeta}$, is a completely symmetric tensor, $\mathscr{K}^{\mu\nu\cdots\zeta} = \mathscr{K}^{(\mu\nu\cdots\zeta)}$, whose symmetrized covariant derivative vanishes. For rank-two Killing tensors in particular, the Killing equation is%
\footnote{The Killing equation arises from the requirement that the Carter constant Poisson commute with the Hamiltonian for geodesic motion, $\mathscr{H}=(1/2)\mathscr{g}^{\mu\nu}p_\mu p_\nu$ (eq. 2.27 of Ref. \cite{Frolov+2017}). This is necessary for any quantity $Q$ (therefore also $E, L, m$) to be conserved along a geodesic since its evolution equation is $\dot{Q} = \{Q, H\}$. The Poisson bracket for phase space observables that are monomials in momenta, e.g., $O = \mathscr{O}^{\alpha\beta\cdots}p_\alpha p_\beta\cdots$, defines an operation called the Nijenhuis-Schouten (NS) bracket on symmetric tensor fields, $\mathbf{\mathscr{O}}$. While the conserved quantities Poisson commute with the Hamiltonian, their underlying Killing tensors NS commute with the metric tensor, leading to eq. \ref{eq:Killing-Equation}.}
\begin{equation} \label{eq:Killing-Equation}
\nabla^{(\alpha}\mathscr{K}^{\mu\nu)} = \mathbf{0}\,.
\end{equation}
Here we have used the parantheses to indicate symmetrization over all enclosed indices, i.e., averaging over all possible permutations, e.g., $S^{(\alpha\mu\nu)} = (S^{\alpha\mu\nu} + S^{\alpha\nu\mu} + S^{\mu\nu\alpha} + S^{\mu\alpha\nu} + S^{\nu\alpha\mu} + S^{\nu\mu\alpha})/3!$. 

The metric tensor itself is clearly a Killing tensor (due to metric compatibility, $\nabla_\alpha \mathscr{g}_{\mu\nu}=\mathbf{0}$). The associated conserved quantity is the particle rest-mass, $m^2 = -\mathscr{g}^{\mu\nu}p_\mu p_\nu$. Below, we will obtain an explicit expression for the Killing tensor associated with the Carter constant, $C$, in the ACKN spacetimes
\begin{equation} \label{eq:Carter-Constant}
C = \mathscr{K}^{\mu\nu}p_\mu p_\nu\,.
\end{equation}

Following the procedure outlined in Sec. 2.2.1 of Ref. \cite{Chervonyi+2015}, we begin by identifying a separation function, $\chi(r, \vartheta) = \chi_r(r) - \chi_\vartheta(\vartheta)$, and two separation tensors, $P^{\mu\nu}(r)$ and $Q^{\mu\nu}(\vartheta)$, which also satisfy $P^{\vartheta \nu} = 0$ and $Q^{r \nu} = 0$, such that the inverse metric tensor can be written as
\begin{equation} \label{eq:Metric-Split}
\mathscr{g}^{\mu\nu} = \frac{P^{\mu\nu} + Q^{\mu\nu}}{\chi}\,.
\end{equation}
Then, the Killing tensor is 
\begin{equation} \label{eq:Killing-Tensor-Formula}
\mathscr{K}^{\mu\nu} = -\frac{\chi_\vartheta P^{\mu\nu} + \chi_r Q^{\mu\nu}}{\chi}\,.
\end{equation}

For the ACKN metric in BL coordinates \eqref{eq:ACKN-Metric}, we identify $\chi(r, \vartheta) = \Sigma(r, \vartheta)$, i.e., $\chi_r =  R^2$ and $\chi_\vartheta = - a^2\cos^2{\vartheta}$. Further, we can find,
\begin{align}
P^{\mu\nu} =&\
-\begin{bmatrix}
\frac{\Pi(\vartheta=\pi/2)}{\Delta} & 0 & 0 & \frac{2aF}{\Delta} \\
0 & -\frac{\Delta}{g} & 0 & 0 \\
0 & 0 & 0 & 0 \\
\frac{2aF}{\Delta} & 0 & 0 & \frac{a^2}{\Delta} \\
\end{bmatrix}\,, \\
Q^{\mu\nu} =&\
+\begin{bmatrix}
-a^2\cos^2{\vartheta} & 0 & 0 & 0 \\
0 & 0 & 0 & 0 \\
0 & 0 & 1 & 0 \\
0 & 0 & 0 & \csc^2{\vartheta}\\
\end{bmatrix}\,.
\end{align}

To match with the ``standard form'' of the Killing tensor for the Kerr metric (see, e.g., eq. 3.22 of Ref. \cite{Frolov+2017}), we change the minus sign in eq. \ref{eq:Killing-Tensor-Formula} and add to it another Killing tensor,
\begin{equation}
\sigma^{\mu\nu} =
\begin{bmatrix}
a^2 & 0 & 0 & a \\
0 & 0 & 0 & 0 \\
0 & 0 & 0 & 0 \\
a & 0 & 0 & 0\\
\end{bmatrix}\,.
\end{equation}
With these modifications, the Killing tensor for the ACKN metric \eqref{eq:ACKN-Metric} is then given as ($s_\vartheta \equiv \sin{\vartheta}$ and $c_\vartheta \equiv \cos{\vartheta}$)
\begin{equation} \label{eq:ACKN-Killing-Tensor}
\mathscr{K}^{\mu\nu} = \frac{a^2c_\vartheta^2}{\Sigma}
\begin{bmatrix}
\frac{(R^2+a^2)2F}{\Delta} & 0 & 0 & \frac{2aF}{\Delta} \\
0 & -\frac{\Delta}{g} & 0 & 0 \\
0 & 0 & \frac{R^2}{a^2c_\vartheta^2} & 0 \\
\frac{2aF}{\Delta} & 0 & 0 & \frac{R^2}{a^2s_\vartheta^2c_\vartheta^2} + \frac{a^2}{\Delta}\\
\end{bmatrix} + \sigma^{\mu\nu}\,.
\end{equation}
We can verify that the above satisfies the Killing equation \eqref{eq:Killing-Equation}.

The massive scalar-wave equation,
\begin{equation} \label{eq:Scalar-Wave-Equation}
(\Box-m^2)\Psi = \frac{1}{\sqrt{-\mathscr{g}}}\partial_\mu\left(\sqrt{-\mathscr{g}}\mathscr{g}^{\mu\nu}\partial_\nu\Psi\right) - m^2 \Psi = 0\,,
\end{equation}
admits a multiplicative separation of variables,
\begin{equation} \label{eq:Scalar-Field}
\Psi(t, r, \vartheta, \varphi) = \mathit{e}^{-\mathit{i}Et}\mathit{e}^{-\mathit{i}m\varphi}\mathscr{R}(r)\Theta(\vartheta)\,,
\end{equation}
if the following complete set of mutually commuting operators, $\{\Box, \mathcal{L}_t, \mathcal{L}_\varphi, \mathcal{K}\}$, exists (see eq. 3.76 of Ref. \cite{Frolov+2017}), where
\begin{align}
\Box =&\ \nabla_\mu\mathscr{g}^{\mu\nu}\nabla_\nu\,,\ \ \
\mathcal{L}_t =\ \mathit{i}(\partial_t)^\mu\nabla_\mu\,, \nonumber \\
\mathcal{K} =&\ \nabla_\mu \mathscr{K}^{\mu\nu}\nabla_\nu\,,\ 
\mathcal{L}_\varphi = \mathit{i}(\partial_\varphi)^\mu\nabla_\mu\,.
\end{align}
In eq. \ref{eq:Scalar-Wave-Equation}, $\mathscr{g}$ denotes the determinant of the metric tensor. The solution of the scalar-wave equation \eqref{eq:Scalar-Field} is the common eigenfunction of the commuting operators. The operators $\mathcal{L}_t$ and $\mathcal{L}_\varphi$ are automatically assured to commute with $\Box$. The separated solution \eqref{eq:Scalar-Field} contains symbols (e.g., $E, \mathscr{R}$) that are analogous to those appearing in the separable geodesic equation \eqref{eq:Geodesic-Tangent}. Since the ACKN metric is identical to the KSZ metric (Sec. \ref{sec:SecIF-Metric-Relations}), which has been shown to admit a separable scalar wave equation, all of these operators must commute (in particular, $\Box$ and $\mathcal{K}$). 

Note that since the Johannsen metric admits a separable geodesic equation, it must possess a Killing tensor \eqref{eq:Carter-Constant}. Further, the complete set of Killing tensors $\{\mathscr{g}^{\mu\nu}, (\partial_t)^\mu, (\partial_\varphi)^\mu, \mathscr{K}^{\mu\nu}\}$ must all Nijenhuis-Schouten commute (see eq. 2.36 of Ref. \cite{Frolov+2017}). However, only a subclass of the Johannsen metric (\textit{viz.}, the ACKN/KSZ metric) has a separable scalar-wave equation and a complete set of commuting Klein-Gordon operators $\{\Box, \mathcal{L}_t, \mathcal{L}_\varphi, \mathcal{K}\}$. We infer from this that $\Box$ and $\mathcal{K}$ do not commute in general for the Johannsen metric. Since the existence of a Killing-Yano tensor (introduced below) ensures that these operators always commute \cite{Carter1977}, we can also infer that the Johannsen metric does not possess a Killing-Yano tensor in general.

Similar to the scalar-wave equation, the separability of the massive Dirac equation requires the existence of a complete set of commuting operators $\{\mathcal{D}, \mathcal{L}_t, \mathcal{L}_\varphi, \mathcal{Y}\}$, whose common eigenfunction is the separated solution. The operator $\mathcal{D}$ is constructed from the Dirac gamma matrices, whereas the operators $\mathcal{L}_t$ and  $\mathcal{L}_\varphi$ also employ the Killing vectors $\partial_t$ and $\partial_\varphi$ in their construction respectively. To the best of our knowledge, the last operator, $\mathcal{Y}$, is always associated with a Killing-Yano tensor. Ref. \cite{Frolov+2017} presents a clear treatment of this topic in the context of the Kerr metric (see eq. 3.86). We now comment on the existence of the Killing-Yano tensor for the ACKN spacetimes.

A Killing-Yano tensor, $\mathscr{Y}_{\mu\nu}$, is the ``square-root'' of the Killing tensor,
\begin{equation} \label{eq:KY-Root-Eq}
\mathscr{K}^{\mu\nu} = -\mathscr{Y}^{\mu\alpha}\mathscr{g}_{\alpha\beta}\mathscr{Y}^{\beta\nu}\,,
\end{equation}
and satisfies the Killing-Yano equation,
\begin{equation} \label{eq:KY-Equation}
\nabla^{(\alpha}\mathscr{Y}^{\mu)\nu} = 0\,.
\end{equation}
Using the above expression for the Killing tensor \eqref{eq:ACKN-Killing-Tensor}, we begin by looking for tensors that satisfy eq. \ref{eq:KY-Root-Eq}, and obtain ($s_\vartheta \equiv \sin{\vartheta}$ and $c_\vartheta \equiv \cos{\vartheta}$)
\begin{equation} \label{eq:ACKN-Killing-Yano}
\mathscr{Y}^{\mu\nu} =
\frac{1}{\Sigma}
\begin{bmatrix}
0 & -a\frac{R^2+a^2}{\sqrt{g}}c_\vartheta & aRs_\vartheta & 0 \\
a\frac{R^2+a^2}{\sqrt{g}}c_\vartheta & 0 & 0 & \frac{a^2}{\sqrt{g}}c_\vartheta \\
-aRs_\vartheta & 0 & 0 & -\frac{R}{s_\vartheta} \\
0 & -\frac{a^2}{\sqrt{g}}c_\vartheta & \frac{R}{s_\vartheta} & 0\\
\end{bmatrix}\,.
\end{equation}
We verify explicitly that the tensor above \eqref{eq:ACKN-Killing-Yano} fails to satisfy the Killing-Yano equation \eqref{eq:KY-Equation} unless $g(r)=1$ and $R(r)=r$. Therefore, a legitimate Killing-Yano tensor exists only for the class of degenerate ACKN spacetimes (see eq. \ref{eq:Degenerate-Seed}). From this, we conclude that the Dirac equation is not separable for non-degenerate ACKN spacetimes. In particular, the Dirac equation is not separable for the class of quasi-degenerate spacetimes, including the Kerr-Sen metric.%
\footnote{The reader may like to see Ref. \cite{Houri+2010} for a discussion on \textit{generalized} Killing-Yano tensors in the Kerr-Sen metric, which lead to the separation of a generalized Dirac equation.} %
We leave the verification of the existence of the complete set of commuting observables for the Dirac equation for the class of degenerate spacetimes for future work but conjecture that they indeed admit a separable Dirac equation.

For the massless limits of the equations discussed above (geodesic, Klein-Gordon, Dirac), it suffices, in general, for the spacetime to possess conformal Killing and conformal Killing-Yano tensors \cite{Frolov+2017}.

Ref. \cite{Collinson1974} discuss how the existence of a Killing-Yano tensor in an empty spacetime implies that the spacetime is of Petrov Type D. We will now show this connection to be true even for non-empty (ACKN) spacetimes in the following section.


\subsection{Petrov Classification of Doubly Separable Spacetimes} 
\label{sec:SecIH-Petrov}

To invariantly classify and distinguish gravitational fields, one can employ the Petrov classification \cite{Petrov2000}. The Petrov classification is based on the algebraic properties of the Weyl tensor—the traceless part of the Riemann tensor that encodes the conformal curvature of spacetime. In vacuum spacetimes, where there is no self-gravitating matter, the Riemann tensor reduces to the Weyl tensor. In four spacetime dimensions, the Weyl tensor generally possesses ten independent components. If two metrics have Weyl tensors of different Petrov types, they cannot be related by a mere change of coordinates—they represent fundamentally different geometries. Following Ref. \cite{Chandrasekhar1985} (see Sec. 9(b)), we employ the Newman-Penrose (NP) formalism \cite{Newman+1961} to identify the Petrov class of the ACKN spacetimes (cf. Ch. 11 of Ref. \cite{Plebanski+2012} for other methods). 

When projected onto an NP tetrad or frame, $\{\ell_+^\mu, \ell_-^\mu, \zeta^\mu, \bar{\zeta}^\mu\}$, which is an orthogonal null tetrad comprised of two real, $\ell_\pm$, and two complex vectors (which are complex conjugates of each other), $\zeta$ and $\bar{\zeta}$, the Weyl tensor $\mathscr{C}_{\beta\mu\nu\gamma}$ is completely encoded into five independent complex scalars,
\begin{align} \label{eq:NP-Weyl-Scalars}
\Psi_0 =&\ \mathscr{C}_{\alpha\beta\gamma\delta}\ell_+^\alpha \zeta^\beta \ell_+^\gamma \zeta^\delta\,, \\
\Psi_1 =&\ \mathscr{C}_{\alpha\beta\gamma\delta}\ell_+^\alpha \ell_-^\beta \ell_+^\gamma \zeta^\delta\,, \nonumber \\
\Psi_2 =&\ \mathscr{C}_{\alpha\beta\gamma\delta}\ell_+^\alpha \zeta^\beta \bar{\zeta}^\gamma \ell_-^\delta\,, \nonumber \\
\Psi_3 =&\ \mathscr{C}_{\alpha\beta\gamma\delta}\ell_+^\alpha \ell_-^\beta \bar{\zeta}^\gamma \ell_-^\delta\,, \nonumber \\
\Psi_4 =&\ \mathscr{C}_{\alpha\beta\gamma\delta}\ell_-^\alpha \bar{\zeta}^\beta \ell_-^\gamma \bar{\zeta}^\delta\,. \nonumber
\end{align}

As interpreted in Sec. 3.5 of Ref. \cite{Szekeres1965} (see also Sec. 3.5 of Ref. \cite{Stephani+2003}), the various Weyl scalars \eqref{eq:NP-Weyl-Scalars} correspond to distinct physical features. A non-zero $\Psi_4$ and a non-zero $\Psi_3$ represent a transverse and longitudinal gravitational wave propagating along $\ell_+$ respectively. A non-zero $\Psi_0$ and a non-zero $\Psi_1$ represent a transverse and longitudinal gravitational wave propagating along $\ell_-$ respectively. Lastly, a non-zero $\Psi_2$ encodes the Coulomb-like or monopolar part of the gravitational field. 

The values of the five complex scalars depends on the choice of the NP tetrad. A null frame can be transformed into another via local Lorentz transformations. Corresponding to the six parameters of the Lorentz group, there are three classes of null rotations. Class I and Class II null rotations leave $\ell_+$ and $\ell_-$ unchanged respectively. Class III null rotations leave the directions of $\ell_\pm$ unchanged and rotate $\zeta$ and $\bar{\zeta}$ by an angle. Explicitly (see Sec. 8(g) of Ref. \cite{Chandrasekhar1985})
\begin{align} \label{eq:Null-Rotations}
\mathrm{Cl.\ I}:&\  \ell_+ \rightarrow \ell_+\,,\ \ell_- \rightarrow c\bar{c}\ell_+ + \ell_- + \bar{c}\zeta + c\bar{\zeta}\,,\ \zeta  \rightarrow c\ell_+ + \zeta \,, \nonumber \\
\mathrm{Cl.\ II}:&\  \ell_+ \rightarrow \ell_+ + b\bar{b}\ell_- + \bar{b}\zeta + b\bar{\zeta}\,,\ \ell_- \rightarrow \ell_-\,,\ \zeta  \rightarrow b\ell_- + \zeta \,, \nonumber \\
\mathrm{Cl.\ III}:&\  \ell_+ \rightarrow n^{-1}\ell_+\,,\ \ell_- \rightarrow n\ell_-\,,\ \zeta  \rightarrow \mathit{e}^{\mathit{i}\omega}\zeta\,.
\end{align}
Here, $c, b, n, \omega$ are the Lorentz group parameters. The first pair are complex whereas the second pair are real. Class I, Class II, and Class III rotations leave $\Psi_0, \Psi_4$ and $\Psi_2$ unchanged respectively. The number and combination of scalars that can be made to vanish by a suitable frame choice define the algebraic Petrov classification of the Weyl tensor.

From a frame in which $\Psi_4 \neq 0$ (always possible to find via a Class I rotation unless all the Weyl scalars vanish), we can perform a Class II null rotation and find the new $0-$Weyl scalar to be related to the old Weyl scalars as
\begin{equation}
\Psi_0^{(\mathrm{new})} = \Psi_4 b^4 + 4\Psi_3 b^3 + 6\Psi_2 b^2 + 4\Psi_1 b + \Psi_0\,.
\end{equation}
By choosing the parameter $b$ appropriately, $\Psi_0^{(\mathrm{new})}$ can be made to vanish, i.e., $b$ has to be a root of the quartic equation
\begin{equation} \label{eq:Petrov-Equation}
\Psi_4 b^4 + 4\Psi_3 b^3 + 6\Psi_2 b^2 + 4\Psi_1 b + \Psi_0 = 0\,.
\end{equation}
A principal null direction (PND) of the Weyl tensor is a direction along which $\Psi_0 = 0$. Therefore, when the original NP frame is chosen such that the first frame leg, $\ell_+$, is a PND, each nonzero root of eq. \ref{eq:Petrov-Equation} produces a new PND \eqref{eq:Null-Rotations}. Each PND satisfies (eq. 382 of Ref. \cite{Chandrasekhar1985})
\begin{equation} \label{eq:PND-Equation}
\Psi_0 = 0 \Leftrightarrow \ell_+^\mu(\ell_+)_{[\alpha}\mathscr{C}_{\beta]\mu\nu[\gamma}(\ell_+)_{\delta]}\ell_+^\nu = 0\,.
\end{equation}
The Petrov classification is then determined by the multiplicity of the roots of the Petrov equation \eqref{eq:Petrov-Equation}. 

When all four roots are distinct, the spacetime is of Petrov Type I. If two of the roots are identical, the spacetime is of Type II. If two pairs of roots are identical, it is of Type D. If exactly three roots coincide, it is of Type III. If all four roots are the same, then the spacetime is of Type N. Finally, if the Weyl tensor itself vanishes, then the spacetime is Type O. 

When the NP frame is chosen such that the first frame leg, $\ell_+$, is a PND, these conditions can be recast into the following conditions on the NP Weyl scalars $\Psi_n$ \cite{Chandrasekhar1985}
\begin{align} \label{eq:Petrov-Type}
\mathrm{Type\ I}:&\ \Psi_0 = 0\,,\\
\mathrm{Type\ II}:&\ \Psi_0 = \Psi_1 = 0\,, \nonumber \\
\mathrm{Type\ D}:&\ \Psi_0 = \Psi_1 = \Psi_3 = \Psi_4 = 0\,, \nonumber \\
\mathrm{Type\ III}:&\ \Psi_0 = \Psi_1 = \Psi_2 = 0\,, \nonumber \\
\mathrm{Type\ N}:&\ \Psi_0 = \Psi_1 = \Psi_2 = \Psi_3 = 0\,, \nonumber \\
\mathrm{Type\ O}:&\ \Psi_0 = \Psi_1 = \Psi_2 = \Psi_3 = \Psi_4 = 0\,. \nonumber 
\end{align}
It should be noted that a Class I null rotation can always be used to additionally set $\Psi_4 = 0$ for the Type I, II and III spacetimes. A Class I null rotation leaves $\Psi_4$ unchanged only when all the other scalars are zero (as in Type N spacetimes).%
\footnote{The expression for the transformed $\Psi_4$ in eq. 342 of Ref. \cite{Chandrasekhar1985} has a typo: In the last term, $\Psi_4$ should be replaced by $\Psi_0$ (see eq. 3.61 of Ref. \cite{Stephani+2003}).}

With all this, we now classify the ACKN spacetimes. We choose the two real null vectors of the NP frame to both be PNDs \eqref{eq:PND-Equation}, which in Boyer-Lindquist coordinates are given as (eq. 2.12 of Ref. \cite{Kocherlakota+2024})
\begin{equation} \label{eq:PNC-Vector-Fields}
\ell_\pm^\mu = \frac{1}{\sqrt{2\Delta\Sigma}}\left[(R^2 + a^2), \pm \frac{\Delta}{\sqrt{g}}, 0, a\right]\,.
\end{equation}
In fact, these PNDs are the generators of the outgoing ($+$) and the ingoing ($-$) principal null congruences of the ACKN spacetimes. We choose the complex vector $\zeta$ to be
\begin{align} \label{eq:NP-Complex-Leg}
\zeta^\mu =&\ \frac{1}{\sqrt{2\Sigma}}\left[\mathit{i}a\sin{\vartheta}, 0, 1, \mathit{i}\csc{\vartheta}\right]\,.
\end{align}
It turns out that the complex tetrad legs also satisfy the PND equation \eqref{eq:PND-Equation}. The tetrad orthonormality relations can be read off from
\begin{equation} \label{eq:NP_Tetrad_Normalization}
\mathscr{g}^{\mu\nu} = -2\ell_+^{(\mu}\ell_-^{\nu)} + 2\zeta^{(\mu}\bar{\zeta}^{\nu)}\,.
\end{equation}

We now compute the NP Weyl scalars \eqref{eq:NP-Weyl-Scalars} explicitly for the ACKN metric in BL coordinates in which $g(r)=1$. The expression for $\Psi_2$ is long and uninsightful.%
\footnote{The $\mathtt{Mathematica}$ code used in this work is available at this repository \cite{Kocherlakota+2025}. 
\label{fn:FN5}} %
Nevertheless, it can be checked that $\Psi_2$ vanishes only for the Minkowski metric. The remaining Weyl scalars are
\begin{align}
&\ \Psi_0 = \Psi_4 = 0\,,\\
&\ \Psi_1 = -\Psi_3 = -\frac{\mathit{i}a\sin{\vartheta}\sqrt{\Delta}}{4\Sigma^3}(R^2 - a^2\sin^2{\vartheta})((\partial_rR)^2-1) \nonumber \\ 
&\ \quad\quad\quad\quad\quad - \frac{\mathit{i}a\sin{\vartheta}\sqrt{\Delta}}{4\Sigma^2}R\partial_r^2R\,, \nonumber
\end{align}
from which it can be seen that $\Psi_1$ and $\Psi_3$ vanish only when $R(r) = r$. Since we already assumed $g(r)=1$, these must be the class of degenerate ACKN metrics (see eq. \eqref{eq:Degenerate-Seed} and below).

From the above, it is evident that none of the ACKN spacetimes can be of Type II since if $\Psi_1$ vanishes then so must $\Psi_3$. Similarly, they cannot be of Type III or N. 
Therefore, the ACKN metrics are in general of Type I. The subclass of degenerate ACKN spacetimes are of Type D. Only the Minkowski metric is of Type O.

The Kerr-Sen metric was discovered to be of Type I and to also admit a separable geodesic equation in Ref. \cite{Hioki+2008}. Here we have shown that a large class of Type I spacetimes also admit a separable geodesic equation. Furthermore, they also admit a separable scalar-wave equation but not a separable Dirac equation (Sec. \ref{sec:SecIG-Killing}).


\section{Self-Gravitating Matter in Doubly Separable Spacetimes} 
\label{sec:SecII-Matter-in-Doubly-Separable-Metrics}

We have discussed above that the ACKN algorithm generates highly symmetric doubly separable spacetimes. The Einstein equations, $\mathscr{G}_{\mu\nu} = 8\pi\mathscr{T}_{\mu\nu}$, where $\mathscr{G}_{\mu\nu}$ is the Einstein tensor%
\footnote{The Einstein tensor, $\mathscr{G}_{\mu\nu}$, is given in terms of the Ricci curvature tensor, $\mathcal{R}_{\mu\nu}$, and its trace, $\mathcal{R} = \mathscr{g}^{\mu\nu}\mathcal{R}_{\mu\nu}$, as $\mathscr{G}_{\mu\nu} = \mathcal{R}_{\mu\nu} - \mathscr{g}_{\mu\nu}\mathcal{R}/2$.} %
and $\mathscr{T}_{\mu\nu}$ the energy-momentum-stress (EMS) tensor of the self-gravitating matter generating the spacetime, must therefore require the matter distribution to obey some symmetry constraints. As an analogy, the spherical symmetry of the hydrogen nucleus underlies the various symmetries of its wavefunction in nonrelativistic quantum mechanics—such as spherical symmetry and separability. In this section, we investigate these symmetry constraints on the matter distribution.

Furthermore, the \textit{type} of the matter field itself imposes constraints on the form of its EMS tensor. We can understand this as follows. The matter EMS tensor in its rest-frame (an orthonomal tetrad $\{e^\mu_{(a)}; a=0\!-\!3\}$ that is adapted to the matter $4-$velocity $e^\mu_{(0)}$) is obtained as
\begin{equation}
\mathscr{T}_{(a)(b)} := \mathscr{T}_{\mu\nu}e^\mu_{(a)}e^\nu_{(b)}\,.    
\end{equation}
The rest-frame EMS tensors for stationary configurations (i.e., no energy or momentum fluxes) of a massless (real) scalar (MS) field, an electromagnetic (EM) field, and a perfect fluid (PF) that are minimally-coupled to gravity must necessarily be of the form (see Appendix F of Ref. \cite{Kocherlakota+2024}):
\begin{align} \label{eq:Matter-Models-RF-EMS-Tensor}
\mathscr{T}_{(a)(b)}^{\mathrm{MS}} =&\ \mathrm{diag.}[\epsilon, +\epsilon, -\epsilon, -\epsilon]\,, \\
\mathscr{T}_{(a)(b)}^{\mathrm{EM}} =&\ \mathrm{diag.}[\epsilon, -\epsilon, +\epsilon, +\epsilon]\,, \nonumber \\
\mathscr{T}_{(a)(b)}^{\mathrm{PF}} =&\ \mathrm{diag.}[\epsilon, +p(\epsilon), +p(\epsilon), +p(\epsilon)]\,. \nonumber
\end{align}
Here we have used $\epsilon$ to denote the energy density of the field and $p(\epsilon)$ to mean the relation between the energy density and the isotropic pressure (or equation of state) of the perfect fluid, both measured in the rest-frame. Therefore, the rest-frame projected Einstein equations, $\mathscr{G}_{(a)(b)} = 8\pi\mathscr{T}_{(a)(b)}$, yield constraints on the components of the Einstein tensor for each specific matter type
\begin{align} 
\mathrm{MS}:&\ +\mathscr{G}_{(r)(r)} = -\mathscr{G}_{(\vartheta)(\vartheta)} = -\mathscr{G}_{(\varphi)(\varphi)} = +\mathscr{G}_{(t)(t)}\,, \\
\mathrm{EM}:&\ -\mathscr{G}_{(r)(r)} = +\mathscr{G}_{(\vartheta)(\vartheta)} = +\mathscr{G}_{(\varphi)(\varphi)} = +\mathscr{G}_{(t)(t)}\,, \nonumber \\
\mathrm{PF}:&\ +\mathscr{G}_{(r)(r)} = +\mathscr{G}_{(\vartheta)(\vartheta)} = +\mathscr{G}_{(\varphi)(\varphi)}\,. \nonumber
\end{align}

Combining the two key ideas — (a) spacetime symmetries constrain the EMS tensor, and (b) choices of matter fields themselves impose structure on the EMS tensor — we are led to ask: 
\begin{quote}
\textit{Can doubly separable spacetimes be sourced by arbitrary types of matter fields?}    
\end{quote}
This question is central to assessing whether the ACKN algorithm can be promoted to a true solution-generating method. If both the seed and generated metrics are sourced by the same type of matter, the EMS tensor must exhibit consistent structure across spin values. Based on this reasoning, we show below that massless scalar fields and perfect fluids cannot support spinning doubly separable spacetimes. An important corollary of this result is that the currently unknown spinning generalization of the Janis-Newman-Winicour naked singularity solution to the Einstein-Klein-Gordon equations does not belong to the class of doubly separable spacetimes. For electromagnetic fields, only the Kerr–Newman family satisfies the separability condition. Physically, this tells us that it is \textit{not} possible to arrange arbitrary matter fields in sufficiently special configurations to create doubly separable spacetimes.

We also noted that spacetimes generated via the AA algorithm feature matter that is rigidly rotating on Boyer–Lindquist (BL) spheres by design, whereas this need not be the case for those generated by the ACKN algorithm. Here, we identify the subset of ACKN-generated separable spacetimes that \textit{do} contain such rigid rotation. This helps clarify which spacetimes are common to both AA and ACKN methods.
Moreover, it is of physical interest to determine whether rigid rotation on spheres is required to ensure equal tangential pressures in the matter rest frame ($p_\vartheta = p_\varphi$).

In Sec. \ref{sec:SecIIA-Matter-Rest-Frame-Review}, we review the construction of the matter rest frame following Ref. \cite{Kocherlakota+2024}. In Sec. \ref{sec:SecIIB-Rigid-Rotation}, we isolate the subclass of separable spacetimes with rigidly rotating matter. In Sec. \ref{sec:SecIIC-Tangential}, we identify a subclass where the tangential pressures are equal, $p_\vartheta = p_\varphi$, or, equivalently, in terms of the Einstein tensor, $\mathscr{G}_{(\vartheta)(\vartheta)} = \mathscr{G}_{(\varphi)(\varphi)}$. This constraint already excludes many spacetimes for which the three matter models in eq. \ref{eq:Matter-Models-RF-EMS-Tensor} would otherwise be considered. In Sec. \ref{sec:SecIID-Meridional}, we find the subclass where the polar and radial pressures satisfy the relations $p_\vartheta = \pm p_r$. Spacetimes that allow $p_\vartheta = p_r$ can contain perfect fluids whereas only those that permit $p_\vartheta = -p_r$ can contain scalar or electromagnetic fields. In Sec. \ref{sec:SecIIE-ACKN-NoGo}, we compile our ``no go'' results, i.e., we identify which of the aforementioned matter fields cannot form doubly separable spacetimes. Sec. \ref{sec:SecIIF-Previous-Results} presents a comparison with previous results.

In Secs. \ref{sec:SecIIB-Rigid-Rotation} and \ref{sec:SecIIC-Tangential}, it is simplest to work in $R=r$ coordinates (see eq. \ref{eq:Seed-Metric-Rr}). This is because the Einstein tensor depends, in addition to the spin parameter $a$ and the polar coordinate $\vartheta$, on the values of the metric functions $f, g, R$, their first-order derivatives $\partial_rf, \partial_rg, \partial_rR$, and the second-order derivatives $\partial_r^2f$ and $\partial_r^2R$. Thus, our coordinate choice guarantees that only a single second-order derivative remains, making downstream analyses significantly more transparent. In Sec. \ref{sec:SecIID-Meridional}, we use coordinates in which $g=1$ (see eq. \ref{eq:Seed-Metric-g1}). This enables us to find a simple closed-form solution to the differential equation involving $R$ and $f$ \eqref{eq:pth-Equal-pr}.


\subsection{Matter Rest Frame in Doubly Separable Spacetimes}
\label{sec:SecIIA-Matter-Rest-Frame-Review}

Since the separable ACKN spacetimes are stationary and axisymmetric, we do not expect any energy or momentum fluxes in the rest frame of the self-gravitating matter. Thus, at each point, $(r,\vartheta)$, there should exist a unique rest frame in which the energy-momentum-stress tensor—or equivalently, the Einstein tensor—is diagonal. Should such a rest frame be absent for a separable spacetime, we are forced to conclude that its self-gravitating matter flows on achronal orbits (it is not a Hawking-Ellis Type-I fluid; see Sec. 4.3 of Ref. \cite{Hawking+1973}). 

The Einstein tensor for the ACKN spacetimes \eqref{eq:ACKN-Metric} has a single off-diagonal component, $\mathscr{G}_{t\varphi}$. Thus, the matter rest frame (where $\mathscr{G}_{(t)(\varphi)}$ vanishes) must be of the form
\begin{align} \label{eq:Rest-Frame}
\begin{alignedat}{3}
& e^\mu_{(t)} =\ \frac{1}{\sqrt{-N_t(\Omega)}}\left[1, 0 ,0, \Omega\right]\,,\
&& e^\mu_{(r)} =\ \frac{1}{\sqrt{+\mathscr{g}_{rr}}}\left[0, 1, 0, 0\right]\,,\\
& e^\mu_{(\varphi)} =\ \frac{1}{\sqrt{+N_\varphi(\psi)}}\left[\psi, 0 ,0, 1\right]\,,\
&& e^\mu_{(\vartheta)} =\ \frac{1}{\sqrt{+\mathscr{g}_{\vartheta\vartheta}}}\left[0, 0, 1, 0\right]\,,
\end{alignedat} 
\end{align}
with the normalizations given as
\begin{align}
N_t(\Omega) =&\ \mathscr{g}_{\mu\nu}e_{(t)}^\mu e_{(t)}^\nu = \mathscr{g}_{tt} + 2\Omega \mathscr{g}_{t\varphi} + \Omega^2\mathscr{g}_{\varphi\varphi}\,, \\
N_\varphi(\psi) =&\ \mathscr{g}_{\mu\nu}e_{(\varphi)}^\mu e_{(\varphi)}^\nu = \psi^2\mathscr{g}_{tt} + 2\psi \mathscr{g}_{t\varphi} + \mathscr{g}_{\varphi\varphi} = \psi^2N_t(\psi^{-1})\,. \nonumber
\end{align}
Requiring that $e^\mu_{(t)}$ and $e^\nu_{(\varphi)}$ be orthonormal ($\mathscr{g}_{\mu\nu}e^\mu_{(t)}e^\nu_{(\varphi)} = 0$) fixes $\psi$ in terms of $\Omega$ as
\begin{align} \label{eq:psi}
\psi(\Omega) =&\ -\frac{\mathscr{g}_{t\varphi} + \Omega\mathscr{g}_{\varphi\varphi}}{\mathscr{g}_{tt} + \Omega\mathscr{g}_{t\varphi}}
= \frac{\mathscr{g}_{\varphi\varphi}(\Omega_{\mathrm{Z}} - \Omega)}{\mathscr{g}_{tt} + \Omega\mathscr{g}_{t\varphi}} \,,
\end{align}
where $\Omega_{\mathrm{Z}} = - \mathscr{g}_{t\varphi}/\mathscr{g}_{\varphi\varphi}$ is the angular velocity of the zero angular momentum observer, i.e., $\mathscr{g}_{\varphi\mu}e_{(t)}^\mu = 0$. Therefore, the problem of finding the matter rest-frame in these spacetimes reduces to finding its angular velocity, $\Omega$. This is done by solving a quadratic equation, $\mathscr{G}_{(t)(\varphi)} = 0$, which is given as
\begin{align} \label{eq:Omega-Quadratic-v0}
& \left[\mathscr{g}_{t\varphi}\mathscr{G}_{\varphi\varphi} - \mathscr{g}_{\varphi\varphi}\mathscr{G}_{t\varphi}\right]\Omega^2 
+ \left[\mathscr{g}_{tt}\mathscr{G}_{\varphi\varphi} - \mathscr{g}_{\varphi\varphi}\mathscr{G}_{tt}\right]\Omega 
\nonumber \\
&\ + \left[\mathscr{g}_{tt}\mathscr{G}_{t\varphi} - \mathscr{g}_{t\varphi}\mathscr{G}_{tt}\right] = 0\,.
\end{align}
For $e_{(t)}$ to be timelike, we require that $N(\Omega) < 0$, i.e., $\Omega_- < \Omega < \Omega_+$, where $\Omega_\pm$ describe null Killing vectors, $N(\Omega_\pm) = 0$. Moreover, if $e_{(t)}$ is timelike, we can be sure that $e_{(\varphi)}$ must be spacelike, i.e., if one of the roots to the equation above corresponds to a timelike vector, then the other must correspond to a spacelike vector. This is because $\psi^{-1}$ satisfies the same equation as $\Omega$ \eqref{eq:Omega-Quadratic-v0}. Thus, spacetimes for which a real root $\Omega$ of eq. \ref{eq:Omega-Quadratic-v0} exists automatically passes the first criterion for physical reasonability, i.e., the matter is assured to flow along timelike orbits. We will denote the legitimate angular velocity of the self-gravitating matter rest-frame by $\Omega_{\mathrm{RF}}(r,\vartheta)$.

The other criterion for physical reasonability is demanding that the self-gravitating matter satisfy some classical energy condition (e.g., the weak energy condition). This can be checked explicitly by employing the matter rest frame energy density, $\epsilon$, and principal pressures, $p_i\ (i=r,\vartheta,\varphi)$, which are given as
\begin{align} \label{eq:Rest_Frame_Matter_Content_Einstein}
\epsilon =\ \mathscr{G}_{\mu\nu}e^\mu_{(t)}e^\nu_{(t)}/(8\pi)\,;\ \ 
p_i =\ \mathscr{G}_{\mu\nu}e^\mu_{(i)}e^\nu_{(i)}/(8\pi)\,. 
\end{align}


\subsection{Doubly Separable Spacetimes with Matter Rigidly Rotating on each Coordinate Sphere}
\label{sec:SecIIB-Rigid-Rotation}

Now that we have discussed how the rest-frame angular velocity of the self-gravitating matter in the ACKN spacetimes can be uniquely obtained, let us identify the class of spacetimes that are common between the AA and ACKN metrics. As noted above, the AA spacetimes contain self-gravitating matter that is rigidly rotating on every coordinate $2-$sphere. Therefore, we find the family of ACKN spacetimes that support matter that is rigidly rotating on spheres.

We start by expressing the quadratic equation for the angular velocity \eqref{eq:Omega-Quadratic-v0} as
\begin{equation} \label{eq:Omega-Quadratic-v1}
\alpha_2(r,y)\Omega^2 + \alpha_1(r,y)\Omega + \alpha_0(r,y) = 0\,.
\end{equation}
Here $y=\cos{\vartheta}$ and the coefficient functions $\alpha_i$ are combinations of components of the metric and Einstein tensors, as given by eq. \ref{eq:Omega-Quadratic-v0}. As noted above, for simplicity, we work in coordinates in which $R(r) = r$ (eq. \ref{eq:Seed-Metric-Rr}). This reduces the functions $\alpha_{i}$ to functions of $f, g, \partial_r f, \partial_r g, \partial_r^2f$ and $y$.

The coefficient functions have the following structure
\begin{align} \label{eq:Coeff_Coeff}
\alpha_2(r,y) =&\ \alpha_{26}(r)y^6 + \alpha_{24}(r)y^4 + \alpha_{22}(r)y^2 + \alpha_{20}(r)\,, 
\nonumber \\
\alpha_1(r,y) =&\ \alpha_{16}(r)y^6 + \alpha_{14}(r)y^4 + \alpha_{12}(r)y^2 + \alpha_{10}(r)\,, 
\nonumber \\
\alpha_0(r,y) =&\ \alpha_{04}(r)y^4 + \alpha_{02}(r)y^2 + \alpha_{00}(r)\,.
\end{align}
The expressions for the new $y-$independent coefficient functions $\alpha_{ij}$ are available in Appendix \ref{AppB:Technical-Details}. Eq. \ref{eq:Coeff_Coeff} allows us to reorganize the quadratic equation for $\Omega$ as
\begin{align} \label{eq:Reorged-Omega-Equation}
&\ [(\alpha_{26}\Omega + \alpha_{16})\Omega] y^6 + [\alpha_{24}\Omega^2 + \alpha_{14}\Omega + \alpha_{04}]y^4
\\
&\ + [\alpha_{22}\Omega^2 + \alpha_{12}\Omega + \alpha_{02}]y^2 + [\alpha_{20}\Omega^2 + \alpha_{10}\Omega + \alpha_{00}] = 0\,. \nonumber 
\end{align}
To find the conditions permitting rigid rotation, it is fruitful to view eq. \ref{eq:Reorged-Omega-Equation} as a cubic equation in $y^2$, i.e., 
\begin{equation} \label{eq:Reorged-Omega-Equation-v2}
\hat{\alpha}_6(r, \Omega) y^6 + \hat{\alpha}_4(r, \Omega) y^4 + \hat{\alpha}_2(r, \Omega) y^2 + \hat{\alpha}_0(r, \Omega) = 0\,.
\end{equation}
The expressions for the new coefficient functions $\hat{\alpha}$ can be read off by comparing eqs. \ref{eq:Reorged-Omega-Equation-v2} and \ref{eq:Reorged-Omega-Equation},
\begin{align}
\hat{\alpha}_6(r, \Omega) =&\ (\alpha_{26}\Omega + \alpha_{16})\Omega\,, \\
\hat{\alpha}_4(r, \Omega) =&\ \alpha_{24}\Omega^2 + \alpha_{14}\Omega + \alpha_{04}\,, \nonumber \\
\hat{\alpha}_2(r, \Omega) =&\ \alpha_{22}\Omega^2 + \alpha_{12}\Omega + \alpha_{02}\,, \nonumber \\
\hat{\alpha}_0(r, \Omega) =&\ \alpha_{20}\Omega^2 + \alpha_{10}\Omega + \alpha_{00}\,. \nonumber
\end{align}
Now, let us suppose that the matter angular velocity is constant on each BL $2-$sphere, i.e., $\Omega_{\mathrm{RF}}(r,\vartheta) = \Omega_\star(r)$. If for some radial coordinate $r$ each coefficient $\hat{\alpha}$ does not vanish, then there are at most only three latitudes on that coordinate $2-$sphere at which $\Omega_\star$ is the legitimate matter angular velocity. Therefore, for separable spacetimes containing rigidly rotating self-gravitating matter, it is necessary that all coefficients $\hat{\alpha}$ vanish identically at each radius $r$. 

On imposing $\hat{\alpha}_6 = 0$, we can see directly from eqs. \ref{eq:Reorged-Omega-Equation-v2} and \ref{eq:Reorged-Omega-Equation} that there are two possibilities,
\begin{itemize}
\item[(I)] $\Omega_\star=0$ and $\alpha_{04} = \alpha_{02} = \alpha_{00} = 0$, or
\item[(II)] $\Omega_\star = -\alpha_{16}/\alpha_{26}$ and $\hat{\alpha}_{4}(\Omega_\star) = \hat{\alpha}_{2}(\Omega_\star) = \hat{\alpha}_{0}(\Omega_\star) = 0$.
\end{itemize}


\subsubsection{Case I}
\label{Sec:SecIIB1-Rigid-Rot-CaseI}

Since the coefficients $\alpha_{0j}$ (see eq. \ref{eq:alpha-0i}) are functions of $f, g, \partial_r f, \partial_r g$ and $\partial_r^2 f$, demanding that each of these vanish yields three coupled ordinary differential equations (ODEs). We obtain two classes of solutions. The first is the trivial class of Kerr spacetimes, which are vacuum spacetimes (no matter fields).
\begin{equation} \label{eq:Rigid-Rotation-Sol1}
\mathrm{Solution\ \#1}:\ R(r) = r\,,\ g(r) = 1\ \mathrm{and}\ f(r) = 1-2M/r\,,
\end{equation}
The second is the class of spacetimes with $f(r) = 1$.
\begin{equation} \label{eq:Rigid-Rotation-Sol1}
\mathrm{Solution\ \#2}:\ R(r) = r\ \mathrm{and}\ f(r) = 1\,.
\end{equation}
The nonspinning seed metric in this case is
\begin{equation} \label{eq:f1_Seed}
\mathrm{d}s^2 = -\mathrm{d}t^2 + g(r)\mathrm{d}r^2 + r^2\mathrm{d}\Omega_2^2\,,
\end{equation}
and the ACKN metric generated from it is
\begin{align} \label{eq:f1_ACKN}
\mathrm{d}s^2 =&\ -\mathrm{d}t^2 + \frac{r^2+a^2\cos^2{\vartheta}}{r^2+a^2}g(r)\mathrm{d}r^2 \\
&\ + (r^2+a^2\cos^2{\vartheta})\mathrm{d}\vartheta^2 + (r^2+a^2)\sin^2{\vartheta}\mathrm{d}\varphi^2\,. \nonumber
\end{align}
For the special case $g(r) = 1$, both metrics (\ref{eq:f1_Seed} and \ref{eq:f1_ACKN}) describe the Minkowski spacetime (in areal and oblate-spheroidal coordinates). When $g(r)\neq 1$, the latter \eqref{eq:f1_ACKN} is a strange nonvacuum spacetime that has no $t\varphi-$component, i.e., it has no frame-dragging ($\mathscr{g}_{t\varphi} = 0$). In this case, the parameter $a$ likely does not represent spin. Since our interest is spinning spacetimes, we do not discuss this metric family further.


\subsubsection{Case II}
\label{Sec:SecIIB1-Rigid-Rot-CaseII}

For this case, the coupled ODEs take the form
\begin{align} \label{eq:ODEs_Rigid_Rotation_CaseII}
0 = \hat{\alpha}_{4}(r, \Omega_\star) =&\ 
(\alpha_{24}\alpha^2_{16} - \alpha_{14}\alpha_{16}\alpha_{26} + \alpha_{04}\alpha^2_{26})/\alpha^2_{26}\,, \nonumber \\
0 = \hat{\alpha}_{2}(r, \Omega_\star) =&\ 
(\alpha_{22}\alpha^2_{16} - \alpha_{12}\alpha_{16}\alpha_{26} + \alpha_{02}\alpha^2_{26})/\alpha^2_{26}\,, \nonumber \\
0 = \hat{\alpha}_{0}(r, \Omega_\star) =&\ 
(\alpha_{20}\alpha^2_{16} - \alpha_{10}\alpha_{16}\alpha_{26} + \alpha_{00}\alpha^2_{26})/\alpha^2_{26}\,,
\end{align}
with the coefficients $\alpha_{ij}$ listed in Appendix \ref{AppB:Technical-Details}. Here as well we obtain two classes of solutions, with the first being the class of Kerr spacetimes. The second is the class of quasi-degenerate spacetimes \eqref{eq:Quasi-Degenerate-Seeds-Rr}.
\begin{equation} \label{eq:Rigid-Rotation}
\mathrm{Solution\ \#3}:\ R(r) = r\ \mathrm{and}\ g(r) = \frac{r^2}{r^2+r_0^2}\,.
\end{equation}
The matter angular velocity in this class of spacetimes is given uniquely, in areal coordinates ($R=r$), by
\begin{equation} \label{eq:Quasi-Degen-Angular-Vel}
\Omega_\star(r) = \frac{a}{r^2 + a^2}\,.
\end{equation}
Therefore, the matter rest-frame in this class of spacetimes is given by the Carter tetrad. To compare, the angular velocity of an equatorial circular timelike geodesic (Keplerian orbit) in these spacetimes is 
\begin{equation}
\Omega_{\rm K;\pm} = \frac{1}{a \pm \sqrt{2r/\partial_rf}}\,,
\end{equation}
where the signs denote prograde ($+$) or retrograde ($-$) rotation. A comparison of the two angular velocities in, for instance, the Kerr–Newman spacetime ($r_0=0, f(r)=1-2M/r+Q^2/r^2$) reveals that the matter rotates with sub-Keplerian angular velocities. This behavior is characteristic (e.g., in Newtonian gravity) of matter fields that possess pressure and exhibit an outwardly decreasing pressure gradient.

The Kerr metric arises as a solution in both cases because the rest-frame quadratic equation becomes fully degenerate: all three coefficient functions, $\alpha_2$, $\alpha_1$, and $\alpha_0$, vanish identically. As a result, any value of $\Omega$ satisfies the equation. This degeneracy is unsurprising, given that the spacetime is vacuum and the stress-energy tensor vanishes everywhere.

Since the quasi-degenerate spacetimes \eqref{eq:Quasi-Degenerate-Seeds-Rr} contain the degenerate spacetimes \eqref{eq:Degenerate-Seed}, the first class of solutions \eqref{eq:Rigid-Rotation-Sol1} is a subset of the third \eqref{eq:Rigid-Rotation}.

As discussed above in Sec. \ref{sec:IIA-Nonspinning-Metrics}, the Kerr-Newman (KN) and the Kerr-Sen (KS) families respectively are important examples of degenerate and quasi-degenerate spacetimes respectively. Therefore, these are also examples of degenerate and quasi-degenerate spacetimes that contain rigidly rotating matter. As a reminder, the KN family is a solution of the Einstein-Maxwell theory \cite{Newman+1965a} whereas the KS family is a solution of the Einstein-Maxwell-dilaton-axion theory that appears in the low-energy effective limit of the heterotic string \cite{Sen1992}. 
 
In our previous work \cite{Kocherlakota+2024} we had shown that quasi-degenerate ACKN spacetimes contain rigidly-rotating matter. Combining the results from both cases discussed above, here we have proved that, in addition to the \textit{nonspinning and axisymmetric} spacetimes with $f(r)=1$ \eqref{eq:f1_ACKN}, the quasi-degenerate spacetimes are the \textit{only} doubly separable spacetimes that contain rigidly-rotating matter. Therefore, these are the only spacetimes that are generated by both the AA and the ACKN algorithms. 


\subsection{Doubly Separable Spacetimes with Equal Tangential Pressures, $p_\vartheta = p_\varphi$}
\label{sec:SecIIC-Tangential}

We have identified the class of separable spacetimes that support rigidly rotating matter. In Ref. \cite{Kocherlakota+2024}, we showed numerically that the class of degenerate metrics have equal tangential pressures, $p_\vartheta = p_\varphi$. But are these the only spacetimes that allow matter with equal tangential pressures in its rest frame? For the same type of matter to source both the seed and the generated ACKN spacetimes, the latter must necessarily admit matter with equal tangential pressures. This crucial requirement stems from the fact that the spherically symmetric seed spacetime inherently supports such a pressure configuration. This is not, in general, necessary if there are multiple self-gravitating matter fields in the spacetime. A notable example of this is the Kerr-Sen solution, which contains a dilaton field, an electromagnetic field, and an axion field, which are also non-mininmally coupled to Einstein-Hilbert gravity (see, e.g., the inset in the right bottom panel of Fig. 4 of Ref. \cite{Kocherlakota+2024}).

We begin by constructing the Einstein tensor as seen by a \textit{generic} timelike Killing observer whose $4-$velocity is given as $k^\mu \propto (1, 0, 0, \Omega)$. To do this, we can still use the tetrad in eq. \ref{eq:Rest-Frame}, without imposing the requirement that $\Omega$ be the true matter angular velocity (i.e., $\Omega \neq \Omega_{\mathrm{RF}}$; See eq. \ref{eq:Omega-Quadratic-v0}). The matter radial and polar pressures take identical values in all such frames, i.e., $p_r = \mathscr{G}^{r}_{\ r}$ and $p_\vartheta = \mathscr{G}^{\vartheta}_{\ \vartheta}$. Only the matter energy-density and azimuthal pressure depend on $\Omega$.

In the frame of the above observer, the tangential pressure difference, $p_\vartheta - p_\varphi(\Omega)$,
is proportional to a quadratic function of $\Omega$, denoted $\Delta p_{\vartheta\varphi}(\Omega)$. Let us express this as
\begin{equation} \label{eq:Tangential_Pressures_Quadratic}
\Delta p_{\vartheta\varphi}(\Omega) = \beta_2(r,y)\Omega^2 + \beta_1(r,y)\Omega + \beta_0(r,y)\,.
\end{equation}
The condition for a separable spacetime to host self-gravitating matter with equal tangential pressures in its rest frame is then simply
\begin{equation} \label{eq:Tangential_Pressures_Quadratic_Eq}
\Delta p_{\vartheta\varphi}(\Omega_{\mathrm{RF}}) = 0\,.
\end{equation}

Thus, we have reduced the current problem to finding a common root, $\Omega_{\mathrm{RF}}$, for the two quadratic equations (\ref{eq:Omega-Quadratic-v1}, \ref{eq:Tangential_Pressures_Quadratic_Eq}),
\begin{align}
\alpha_2 \Omega^2 + \alpha_1 \Omega + \alpha_0 = 0\ \mathrm{and}\
\beta_2 \Omega^2 + \beta_1 \Omega + \beta_0 = 0\,.
\end{align}
A common root exists when the resultant of the two quadratics, $\Gamma$, vanishes. The resultant is given in terms of the coefficients of the two quadratics as
\begin{equation} \label{eq:Common_Root}
\Gamma = (\alpha_2\beta_0 - \alpha_0\beta_2)^2 - (\alpha_1\beta_0 - \alpha_0\beta_1)(\alpha_2\beta_1 - \alpha_1\beta_2)\,.
\end{equation}
Here, it has the following structure 
\begin{equation} \label{eq:Tenth_Order}
\Gamma(r,y) = -4a^2 \Sigma^4(r,y)\Delta(r)\Gamma_2(r,y)\Gamma_4(r,y)\,,
\end{equation}
where $\Gamma_2$ and $\Gamma_4$ are a quadratic and a quartic polynomial in $y^2$ respectively,
\begin{align} \label{eq:Quadratic-Quartic}
\Gamma_2(r,y) =&\ \gamma_{24}(r)y^4 + \gamma_{22}(r)y^2 + \gamma_{20}(r)\,, \\
\Gamma_4(r,y) =&\ \gamma_{48}(r)y^8 + \gamma_{46}(r)y^6 + \gamma_{44}(r)y^4 \nonumber \\
&\ + \gamma_{42}(r)y^2 + \gamma_{40}(r)\,. \nonumber
\end{align}
Since we are working in coordinates in which $R(r)=r$, the coefficient functions $\gamma_{ij}$ are combinations of the metric functions, $f$ and $g$, and their derivatives, $\partial_r f, \partial_r g$ and $\partial^2_r f$ (for further details, see Appendix \ref{AppB:Technical-Details}). 

The resultant vanishes when either $\Gamma_2$ or $\Gamma_4$ vanishes ($\Delta$ vanishes only at a horizon). For the case when $\Gamma_2=0$, if all the coefficients $\gamma_{2i}$ do not vanish at some radius, then the matter has equal tangential pressures in its rest-frame only at specific latitudes on that $2-$sphere. Therefore, for our purposes, we require that each coefficient $\gamma_{2i}$ vanish identically. This gives us three equations in five unknowns ($f, g, \partial_r f, \partial_r g, \partial^2_r f$). Similarly, for the case when $\Gamma_4=0$, we need all coefficients $\gamma_{4i}$ to vanish. Here we obtain five equations in five unknowns. 

The only solution we obtain from requiring that the coefficients $\gamma_{2i}$ vanish is the class of degenerate spacetimes \eqref{eq:Degenerate-Seed}.
\begin{equation} \label{eq:Equal-Tangential-Sol1}
\mathrm{Solution\ \#1}:\ R(r)=r\ \mathrm{and}\ g(r)=1.
\end{equation}
By requiring that the coefficients $\gamma_{4i}$ vanish, we get two solutions. The first is the Kerr metric, which is a trivial solution since, being vacuum, it has $p_\vartheta = p_\varphi = 0$. The second is a subfamily of the nonspinning and axisymmetric $f(r)=1$ spacetimes discussed above in Sec. \ref{Sec:SecIIB1-Rigid-Rot-CaseI}.
\begin{equation} \label{eq:Equal-Tangential-Sol2}
\mathrm{Solution\ \#2}:\ R(r)=r\,,\ g(r) = \frac{r^2}{r^2+r_0^2}\ \mathrm{and}\ f(r)=1\,,    
\end{equation}
where $r_0$ is a constant. There is a specific (null) frame in which the tangential pressures become equal.

In summary, we find that the only spinning spacetimes of the ACKN family of mertrics with equal tangential pressures are those derived from the degenerate seeds \eqref{eq:Degenerate-Seed}.


\subsection{Doubly Separable Spacetimes with Equal Meridional Pressures, $|p_\vartheta| = |p_r|$}
\label{sec:SecIID-Meridional}

We have seen above \eqref{eq:Matter-Models-RF-EMS-Tensor} that perfect fluids, scalar fields and electromagnetic fields have equal meridional pressure magnitudes, $|p_\vartheta| = |p_r|$. More specifically, for perfect fluids these are related as $p_\vartheta = p_r$, whereas for the scalar and electromagnetic fields, they are given as $p_\vartheta = -p_r$. Here we will identify the families of doubly separable spacetimes that can potentially host these different types of matter.

As noted above, the rest-frame expressions for $p_\vartheta$ and $p_r$ are independent of the matter angular velocity $\Omega_{\mathrm{RF}}$. Their expressions have the following forms
\begin{align} \label{eq:Meridional-Pressures}
p_\vartheta =&\ \frac{p_{\vartheta 4}(r) y^4 + p_{\vartheta 2}(r) y^2 + p_{\vartheta 0}(r)}{32\pi\Sigma^3 g^2}\,, \\
p_r =&\ \frac{p_{r 2}(r) y^2 + p_{r 0}(r)}{32\pi\Sigma^3 g^2}\,. \nonumber
\end{align}
The coefficient functions, $p_{\vartheta i}$ and $p_{r i}$, are displayed in Appendix \ref{AppB:Technical-Details}.

For the meridional pressures to be equal in magnitude, i.e., $p_\vartheta = \pm p_r$, it is necessary that the following quadratic equations in $y$ are satisfied,
\begin{equation}
p_{\vartheta 4}(r)y^4 + [p_{\vartheta 2}(r) \mp p_{r 2}(r)]y^2 + [p_{\vartheta 0}(r) \mp p_{r 0}(r)] = 0\,.
\end{equation}
For them to be satisfied for all $y$, we require once again that each coefficient vanish identically. 

It can be shown straightforwardly then that $p_\vartheta = +p_r$ is possible only in the Kerr metric (where $p_\vartheta$ and $p_r$ are both zero):
\begin{equation} \label{eq:pth-Equal-pr}
p_\vartheta = +p_r:\ R(r) = r,\ g(r) = 1\ \mathrm{and}\ f(r) = 1-2M/r\,.
\end{equation}
This is a vacuum metric. We can thus rule out the possibility that perfect fluids can be matter sources of nontrivial spinning doubly separable spacetimes. 

The equation $p_\vartheta = -p_r$ turns out to be independent of $y$. Therefore, we obtain a single ODE involving the metric functions and their derivatives. Finding a solution to this ODE is most convenient in coordinates in which $g(r) = 1$. Therefore, this ODE involves $f$ and $R$, their first-order derivatives as well as their second-order derivatives. This ODE turns out to be integrable and we find the following solution for $f$ in terms of $R$ and two integration constants, $\alpha$ and $\beta$,
\begin{equation} \label{eq:pth-Equal-pr}
p_\vartheta = -p_r:\ g(r) = 1\ \mathrm{and}\ f(r) = \frac{r^2 + \alpha r + \beta}{R^2(r)}\,.
\end{equation}
It can be seen immediately that the Kerr-Newman (KN) solution belongs to this metric family. For a KN spacetime of mass $M$ and charge $Q$, we can identify that
\begin{equation} \label{eq:KN}
R(r)=r\,,\ \alpha = -2M\,,\ \beta = + Q^2\,.
\end{equation}
Note that this is not the only possible choice, i.e., eq. \ref{eq:pth-Equal-pr} does not imply the KN metric uniquely (e.g., $R(r) \neq r$).

Table \ref{table:Matter_Prop} compiles our results from Sec. \ref{sec:SecIIB-Rigid-Rotation}, \ref{sec:SecIIC-Tangential}, and Sec. \ref{sec:SecIID-Meridional}.


\begin{table*}[t]
\begin{center}
\caption{\textit{Summary of how specific physical conditions of the self-gravitating matter are permitted only in specific classes of doubly separable ACKN/KSZ spacetimes. Here, $\Omega_{\rm RF}$ denotes the matter rest frame angular velocity in these spacetimes.}}
\label{table:Matter_Prop}
\renewcommand{\arraystretch}{1.5}
\centering
\begin{tabular}[t]{
!{\vrule width 1.5pt}
c
!{\vrule width 1.5pt}
c
!{\vrule width 1.5pt}
c
!{\vrule width 1pt}
c
!{\vrule width 1pt}
c
!{\vrule width 1pt}
c
!{\vrule width 1.5pt}
c
!{\vrule width 1.5pt}
c
!{\vrule width 1.5pt}
}
\Xhline{1.2pt}  
Physical Condition & Equation & $\#$ & $R(r)$ & $g(r)$ & $f(r)$ & $\Omega_{\mathrm{RF}}$ & Metric Class\\
\Xhline{1.2pt}  
Rigid Rotation & $\Omega_{\mathrm{RF}} = \Omega_\star(r)$ & 1. & $r$ & $\frac{r^2}{r^2+r_0^2}$ & any $f$ & $\frac{a}{r^2 + a^2}$ & Quasi-Degenerate \\
 & & & $\sqrt{(r+r_1)^2 - r_0^2}$ & $1$ & any $f$ & $\frac{a}{R^2 + a^2}$ & \\
\cline{3-8}
& & 2. & $r$ & any $g$ & $1$ & any $\Omega_{\rm RF}$ & Nonspinning and Axisymmetric\\
\Xhline{1.2pt} 
Equal Tangential Pressures & $p_\vartheta = p_\varphi$ & 3. & $r$ & $1$ & any $f$ & $\frac{a}{r^2 + a^2}$ & Degenerate \\
\cline{3-8}
& & 4. & $r$ & $\frac{r^2}{r^2+r_0^2}$ & $1$ & $\pm \frac{\csc{\vartheta}}{\sqrt{r^2+a^2}}$ & Nonspinning and Axisymmetric\\
\Xhline{1.2pt} 
Equal Meridional Pressures & $p_\vartheta = +p_r$ & 5. & $r$ & $1$ & $1-2M/r$ & any $\Omega_{\rm RF}$ & Kerr \\
\cline{2-8}
& $p_\vartheta = -p_r$ & 6. & any $R$ & $1$ & $\frac{r^2 + \alpha r + \beta}{R^2}$ & $\Omega_{\rm RF}$ from eq. \ref{eq:Omega-Quadratic-v0} & Contains Kerr-Newman \\
\Xhline{1.2pt} 
\end{tabular}
\end{center}
\end{table*}


\subsection{Doubly Separable Spacetimes and Possible Matter Sources: Perfect Fluids, Massless Scalar Fields and Electromagnetic Fields}
\label{sec:SecIIE-ACKN-NoGo}

We have now established various components of the criteria \eqref{eq:Matter-Models-RF-EMS-Tensor} needed to test whether doubly separable spacetimes can be sourced by our three fiducial matter models, \textit{viz.}, perfect fluids, scalar fields, and electromagnetic fields.

Perfect fluids require $p_\vartheta = p_\varphi = +p_r$. We saw how the requirement $p_\vartheta = p_r$ rules out perfect fluids as sources of doubly separable spacetimes, except the trivial Kerr metric when the pressure is zero \eqref{eq:pth-Equal-pr}. Therefore, it is not possible for the spacetimes of generic neutron stars, which are composed of perfect fluids, to exhibit high degrees of symmetry, i.e., they cannot be doubly separable spacetimes. 

On the other hand, we can see from eq. \ref{eq:Matter-Models-RF-EMS-Tensor} that both massless real scalar fields and electromagnetic fields require $p_\vartheta = p_\varphi = -p_r$. Combining the results from Secs. \ref{sec:SecIIC-Tangential} and \ref{sec:SecIID-Meridional}, we see that the only doubly separable spacetimes that can contain such matter models are those with (eqs. \ref{eq:pth-Equal-pr}, \ref{eq:Equal-Tangential-Sol1})
\begin{equation}
f(r) = \frac{r^2 + \alpha r + \beta}{R^2(r)}\,, g(r) = 1\ \mathrm{and}\ R(r) = r\,,
\end{equation}
or, equivalently,
\begin{equation}
f(r) = 1 + \frac{\alpha}{r} + \frac{\beta}{r^2}\,,\ g(r) = 1\ \mathrm{and}\ R(r) = r\,.
\end{equation}
For the above class of spacetimes, the unique rest frame angular velocity is given by \eqref{eq:Quasi-Degen-Angular-Vel}, using which we can show that the EMS tensor has the following form
\begin{equation}
\epsilon = \frac{\beta}{8\pi\Sigma^2}\,;\ 1 = -\frac{p_r}{\epsilon} = +\frac{p_\vartheta}{\epsilon} = +\frac{p_\varphi}{\epsilon}\,.
\end{equation}
Therefore, this class of spacetimes contains an electromagnetic field (since $p_r = -\epsilon$). Indeed, these are the Kerr-Newman spacetimes \eqref{eq:KN}. Clearly, the only doubly separable spacetimes that contain an electromagnetic field are the KN spacetimes.

Furthermore, there are no doubly separable spacetimes that can be sourced by a massless real scalar field. In particular, the spinning generalization of the Janis-Newman-Winicour (JNW) naked singularity solution cannot belong to the class of doubly separable spacetimes.


\subsection{Comparison with Previous Results}
\label{sec:SecIIF-Previous-Results}

\subsubsection{Spinning Janis-Newman-Winicour Spacetimes}

A spinning generalization of the JNW solution was proposed in Ref. \cite{Solanki+2022} using the AA algorithm. The authors implicitly adopt a choice for the conformal factor, \textit{viz.}, $\hat{X}=\hat{\Sigma}$ (see the discussion in Sec. \ref{sec:SecIC-AA-Metric}). Furthermore, the matter rest-frame is assumed to be given by the Carter tetrad, i.e., by the tetrad in eq. \ref{eq:Rest-Frame} with $\Omega$ given by eq. \ref{eq:Quasi-Degen-Angular-Vel}. We now argue that these choices are not appropriate and do not lead to a spinning spacetime generated by a scalar field.

The metric for the nonspinning JNW seed metric is (see eq. 1 of Ref. \cite{Solanki+2022})
\begin{align} \label{eq:JNW-Metric}
\mathrm{d}s^2 =&\ -\left(1-\frac{2M}{r\nu}\right)^\nu\mathrm{d}t^2 + \left(1-\frac{2M}{r\nu}\right)^{-\nu}\mathrm{d}r^2 \nonumber\\
&\ + r^2\left(1-\frac{2M}{r\nu}\right)^{1-\nu}\mathrm{d}\Omega_2^2\,,
\end{align}
where $M$ denotes the total (Arnowitt-Deser-Misner; \cite{Arnowitt+2008}) mass of the spacetime and $\nu$ is the scalar ``charge,'' which controls the amplitude of the scalar field (eq. 12 of Ref. \cite{Virbhadra1997}),
\begin{equation} \label{eq:JNW-Scalar-Field}
\Phi(r) = \frac{\sqrt{1-\nu^2}}{\sqrt{16\pi}}\left(1-\frac{2M}{r\nu}\right)\,.
\end{equation}
The metric tensor \eqref{eq:JNW-Metric} and the scalar field \eqref{eq:JNW-Scalar-Field} together produce a legitimate solution of the Einstein equations as well as the massless Klein-Gordon equation.

When a seed metric is written in $g=1$ coordinates \eqref{eq:Seed-Metric-g1}, for the aforementioned choice of the conformal factor ($\hat{X}=\hat{\Sigma}$), the spinning spacetimes generated by the AA and ACKN algorithms are identical (compare eqs. \ref{eq:AA-Metric} and \ref{eq:ACKN-Metric}). Notice that the JNW seed is expressed in such coordinates in eq. \ref{eq:JNW-Metric}.

Furthermore, it is apparent that the JNW seed metric does not belong to the class of quasi-degenerate spacetimes \eqref{eq:Quasi-Degenerate-Seeds-g1}. Therefore, the self-gravitating matter in the spinning ACKN spacetime cannot be rigidly-rotating on Boyer-Lindquist $2-$spheres (Sec. \ref{sec:SecIIB-Rigid-Rotation}). That is, the matter rest-frame cannot be given by the Carter frame. The matter stress tensor associated with the spinning JNW metric in Ref. \cite{Solanki+2022} (eq. 46) will have a non-zero $t\varphi-$component in the Carter frame. Equivalently, the second partial differential equation in eq. \ref{eq:AA_PDEs}, required to fix the conformal factor in the AA algorithm, is not satisfied. This is consistent with the discussion in the previous section: The spinning JNW metric cannot belong to the class of ACKN spacetimes. Equivalently, it cannot belong to the class of AA spacetimes with $\hat{X}=\hat{\Sigma}$ (in $g=1$ coordinates). 

Following the discussion in Sec. \ref{sec:SecIIE-ACKN-NoGo} and above, we conclude that the recently discovered legitimate spinning generalization of the JNW solution, \textit{viz.}, the Mirza–Kangazi–Sadeghi solution \cite{Mirza+2023}, cannot belong to the class of ACKN spacetimes.

\subsubsection{No-Go Theorems for Electromagnetic Fields and Perfect Fluids}

In Ref. \cite{Drake+2000}, the authors--motivated by considerations similar to our own--investigate under what conditions the Drake–Szekeres (DS) metric can admit electromagnetic fields or perfect fluid matter content. They establish two key results: (1) the only DS spacetime that supports an electromagnetic field is the Kerr–Newman (KN) solution, and (2) the only DS spacetime compatible with a perfect fluid is the vacuum. These conclusions are in complete agreement with our findings.

Since the ACKN family constitutes a subclass of the DS metrics, it is natural to expect that our results are encompassed within those of Ref. \cite{Drake+2000}. However, a noteworthy distinction lies in the assumptions for the electromagnetic case: while we adopt a different criterion (that the electromagnetic stress tensor should take the form in eq. \ref{eq:Matter-Models-RF-EMS-Tensor} in its rest frame), Ref. \cite{Drake+2000} impose the vanishing of the Ricci scalar—a necessary condition for solutions to the Einstein–Maxwell equations—and show that this requirement uniquely selects the Kerr–Newman spacetime within the DS class. Our result may thus be regarded as a corollary of theirs upon restriction to the ACKN subclass.

In the case of perfect fluids, however, we find the argument presented in Ref. \cite{Drake+2000}—which aims to exclude the possibility of nontrivial DS spacetimes containing self-gravitating perfect fluids—to be lacking. The Einstein tensor for a generic DS spacetime includes two off-diagonal components: $t\varphi$, as in the ACKN metric, and also $r\vartheta$. The authors assume the vanishing of the $r\vartheta-$component in their proof, a requirement that lacks sufficient justification. In spacetimes where the Einstein tensor has two off-diagonal components, the matter rest frame must be adapted from that of eq. \ref{eq:Rest-Frame} by incorporating revised legs along the $r$ and $\vartheta$ directions,
\begin{align} \label{eq:Rest-Frame-ry}
\begin{alignedat}{3}
e^\mu_{(r)} \propto\ \left[0, 1, x_r, 0\right]\,;\ 
e^\mu_{(\vartheta)} \propto\ \left[0, x_\vartheta, 1, 0\right]\,,
\end{alignedat} 
\end{align}
which satisfy the orthonormalization conditions $\mathscr{g}_{\mu\nu}e^\mu_{(r)}e^\nu_{(\vartheta)} = 0$ (this fixes $x_\vartheta$ in terms of $x_r$) and cause the \textit{projected} $r\vartheta-$component of the Einstein tensor, $\mathscr{G}_{(r)(\vartheta)} = \mathscr{G}_{\mu\nu}e^\mu_{(r)}e^\nu_{(\vartheta)}$, to vanish (this fixes $x_r$). It is the projection of the Einstein tensor in this frame that must be analyzed to determine whether the Kerr metric is indeed the only DS spacetime that can support a perfect fluid.


\section{Conclusions}
\label{sec:SecIV-Conclusions}

In this work, we have clarified the structure and physical relevance of a class of spacetimes generated by the Azreg-Aïnou–Chen–Kocherlakota–Narayan (ACKN) algorithm. We demonstrated that this construction yields the Konoplya-Stuchlík-Zhidenko (KSZ) metric, which is both geodesically and scalar-wave separable—thereby deserving the label ``doubly separable.'' Within this class, we identified a subclass of \textit{degenerate} ACKN spacetimes that also admit a Killing-Yano tensor and hence might support a separable Dirac equation. We showed that the latter sub-class is Type D in the Petrov classification, while the general ACKN spacetime is Type I. The rich symmetry structures of the ACKN spacetimes make them highly useful for analytical studies of particle dynamics, wave propagation, and quantum field behavior in curved spacetime.

We have uncovered a sharp interplay between the dynamics of the self-gravitating matter and the hidden symmetries of the spacetime. Rigid rotation on Boyer–Lindquist 2-spheres is possible only in \textit{quasi-degenerate} ACKN spacetimes (a one-parameter extension of the degenerate class) while isotropic tangential pressures ($p_\vartheta = p_\varphi$) are admitted exclusively by the fully degenerate ACKN spacetimes. The class of ACKN spacetimes that admit matter with isotropic meridional pressures is identified in Sec. \ref{sec:SecIID-Meridional}.

More fundamentally, our analysis demonstrates that the existence of such symmetries places strong restrictions on the allowable forms of self-gravitating matter. This stems from the fact that different matter fields obey distinct equations of state, i.e., fixed relationships between their pressure and energy density. In particular, we found that perfect fluids and massless real scalar fields cannot source doubly separable spacetimes of this kind, while electromagnetic fields yield only the well-known Kerr-Newman family. This effectively demonstrates that the legitimate spinning generalization of the Janis-Newman-Winicour solution, \textit{viz.}, the Mirza–Kangazi–Sadeghi solution \cite{Mirza+2023}, is not a doubly separable spacetime.

Our results not only clarify the scope and limitations of a widely used solution-generating algorithm but also emphasize the importance of compatibility between spacetime symmetries and matter field configurations. In this light, the ACKN framework provides a powerful tool for constructing viable spinning spacetimes, particularly for use in phenomenological modeling. At the same time, our findings highlight the need for caution: metric ansatzes derived through symmetry-based algorithms must be scrutinized for consistency with the full Einstein-matter equations.

Although our results apply beyond black hole spacetimes, their relevance to black holes is confined to their exterior. Extending our analysis to include the interior is straightforward.

Looking forward, the methods developed here could be applied to a wider variety of matter models, including vector and tensor fields or more exotic configurations. They may also help in identifying interior solutions matching to known exteriors or in refining the criteria for ultracompact object formation and stability in general relativity.


\begin{acknowledgments}
We thank Behrouz Mirza for bringing their spinning JNW naked singularity solution to our attention, sharing their code, and helpful discussions.

Support comes from grants from the Gordon and Betty Moore Foundation (GBMF-8273) and the John Templeton Foundation (\#62286) to the Black Hole Initiative at Harvard University. 

The data that support the findings of this article are openly available at this repository \cite{Kocherlakota+2025}. \textit{Software:} $\mathtt{Matplotlib}$ \cite{Hunter2007}, $\mathtt{diffgeo.m}$ \cite{Headrick2024}.
\end{acknowledgments}

\providecommand{\noopsort}[1]{}\providecommand{\singleletter}[1]{#1}%

\begin{appendix}


\section{Relating Boyer-Lindquist Coordinate Systems}
\label{AppA:BL-Systems}

Since we are interested in Sec. \ref{sec:SecI-Separable-Spacetimes} in the relationship between different stationary and axisymmetric metric ansatzes, all written in Boyer-Lindquist (BL; Ref. \cite{Boyer+1967}) coordinates, here we will describe how the form of any particular stationary and axisymmetric spacetime metric depends on the choice of the BL coordinates. 

Consider a spacetime metric, $\mathscr{g}_{\hat{\mu}\hat{\nu}}$, written in an arbitrary set of coordinates, $x^{\hat{\mu}} = (T, \rho, \psi, \phi)$. Upon performing a coordinate transformation, $x^{\hat{\mu}} \mapsto x^\mu = (t, r, \vartheta, \varphi)$, the Jacobian, $\Lambda^\mu_{\ \hat{\mu}}$, of the transformation relates the basis one-forms of the two systems as
\begin{equation}
\mathrm{d}x^\mu = \Lambda^\mu_{\ \hat{\mu}}\mathrm{d}x^{\hat{\mu}}\,.
\end{equation}
The basis vectors are related as
\begin{equation}
\partial_\mu = \Lambda_\mu^{\ \hat{\mu}}\partial_{\hat{\mu}}\,,
\end{equation}
and the metric tensors are related via a similarity transformation,
\begin{equation}
\mathscr{g}_{\mu\nu} = \Lambda_\mu^{\ \hat{\mu}}\mathscr{g}_{\hat{\mu}\hat{\nu}}\Lambda_{\ \nu}^{\hat{\nu}}\,.
\end{equation}
In matrix notation, the above can be expressed as $\mathrm{d} = \Lambda \cdot \hat{\mathrm{d}}$, $\partial = (\Lambda^{-1})^{\mathrm{T}} \cdot \hat{\partial}$, and $\mathscr{g} = (\Lambda^{-1})^{\mathrm{T}} \cdot \hat{\mathscr{g}} \cdot \Lambda^{-1}$ respectively.

Schwarz's theorem on the equality of mixed partials implies that the commutators of each set of basis vectors must vanish identically, i.e., 
\begin{equation} \label{eq:Equality-Mixed-Partials}
[\partial_{\hat{\mu}}, \partial_{\hat{\nu}}] = \mathbf{0}\,,\ [\partial_\mu, \partial_\nu] = \mathbf{0}\,,
\end{equation}
where the commutator $[\mathbf{v}, \mathbf{w}]^a$ of two arbitrary vectors $\mathbf{v} = v^a \partial_a$ and $\mathbf{w} = w^a\partial_a$ is defined as $[\mathbf{v}, \mathbf{w}]^a := v^b\partial_b w^a - w^b\partial_b v^a$. This requirement \eqref{eq:Equality-Mixed-Partials} imposes constraints on the Jacobian (see also Problems 3-5 of Ch. 1 of Ref. \cite{Wald1984} and Sec. 6.2 of Ref. \cite{Plebanski+2012}),
\begin{align} \label{eq:Vanishing-Structure-Constants}
0 =&\ [\partial_\mu, \partial_\nu]^\alpha = [\Lambda_\mu^{\ \hat{\mu}}\partial_{\hat{\mu}}, \Lambda_\nu^{\ \hat{\nu}}\partial_{\hat{\nu}}]^{\hat{\alpha}}\,,\quad \mathrm{i.e.},
\nonumber\\
0 =&\ \Lambda_\mu^{\ \hat{\mu}}\partial_{\hat{\mu}}\Lambda_\nu^{\ \hat{\alpha}} - \Lambda_\nu^{\ \hat{\nu}}\partial_{\hat{\nu}}\Lambda_\mu^{\ \hat{\alpha}} 
= 2\Lambda_{[\mu}^{\ \ \hat{\beta}}\Lambda_{\nu], \hat{\beta}}^{\ \ \hat{\alpha}} =: \mathscr{C}^{\hat{\alpha}}_{\mu\nu}\,. 
\end{align}
In the above, we have introduced the structure constants, $\mathscr{C}^{\hat{\alpha}}_{\mu\nu}$, of the new basis vector fields $\{\partial_\mu\}$ in the old coordinates $x^{\hat{\alpha}}$ (see, e.g., eq. 7.2.5 of Ref. \cite{Wald1984}).

Let us now concern ourselves specifically with transformations linking two BL systems. Let $x^{\hat{\mu}}$ and $x^\mu$ be the old and new set of BL coordinates respectively. To prevent mixing between the $r\vartheta-$sector and the $t\varphi-$sector of the metric, we will require that $t=t(T, \phi), r=r(\rho, \psi), \vartheta=\vartheta(\rho, \psi)$, and $\varphi=\varphi(T, \phi)$. This ensures that the $tr, t\varphi, r\varphi$, and $\vartheta\varphi-$components of the metric vanish. For $\mathscr{g}_{r\vartheta} = 0$, we require that 
\begin{equation}
\mathscr{g}_{\rho\rho}(\partial_{\rho}r)(\partial_\psi r) + \mathscr{g}_{\psi\psi}(\partial_{\rho}\vartheta)(\partial_\psi \vartheta) = 0\,.
\end{equation}

For convenience, we will choose that $\partial_\psi r = 0 = \partial_{\rho}\vartheta$. Thus, the Jacobian is of the form,
\begin{equation} \label{eq:BL-Jac}
\Lambda^\mu_{\ \hat{\mu}} = 
\begin{bmatrix}
\alpha & 0 & 0 & \beta \\
0 & \chi & 0 & 0 \\
0 & 0 & \xi & 0 \\
\gamma & 0 & 0 & \delta
\end{bmatrix}\,,
\end{equation}
where $\alpha = \partial_T t, \beta = \partial_\phi t, \gamma = \partial_T \varphi,$ and $\delta = \partial_\phi \varphi$ are functions of $T$ and $\phi$. Furthermore, we have introduced $\chi$ and $\xi$ above to denote $\chi := \partial_\rho r$ and $\xi := \partial_\psi \vartheta$.

The requirement that the structure constants vanish \eqref{eq:Vanishing-Structure-Constants} for the Jacobian above \eqref{eq:BL-Jac} yields ten differential equations, eight of which are 
$([\partial_t, \partial_r] =\mathbf{0}, [\partial_t, \partial_\vartheta]=\mathbf{0}, ([\partial_\varphi, \partial_r] =\mathbf{0}, [\partial_\varphi, \partial_\vartheta]=\mathbf{0})$
\begin{equation} \label{eq:DEs-1}
\partial_\rho[\lambda\zeta] = 0\,;\ \partial_\psi[\lambda\zeta] = 0\,,
\end{equation}
where $\lambda := (\alpha\delta - \beta\gamma)^{-1}$ and $\zeta$ can be either $\alpha, \beta, \gamma$ or $\delta$. These are trivially satisfied since none of these functions depend on $\rho$ or $\psi$. The remaining two equations are $([\partial_t, \partial_\varphi]=\mathbf{0})$
\begin{align} \label{eq:DEs-2}
-\delta\partial_T(\lambda \beta) + \beta\partial_T(\lambda\delta) 
- \alpha\partial_\phi(\lambda\delta) + \gamma\partial_\phi(\lambda\beta) &=\ 0\,,\\
+\delta\partial_T(\lambda \alpha) - \beta\partial_T(\lambda\gamma) 
+ \alpha\partial_\phi(\lambda\gamma) - \gamma\partial_\phi(\lambda\alpha) &=\ 0\,. \nonumber
\end{align}

Before attempting to solve the equations above \eqref{eq:DEs-2}, we note that the metric functions in the new coordinates are
\begin{align} \label{eq:New_BL_Metric}
\mathscr{g}_{tt} =&\ \lambda^2(\delta^2 \mathscr{g}_{TT} - 2\gamma\delta\mathscr{g}_{T\phi} + \gamma^2\mathscr{g}_{\phi\phi})\,, \\
\mathscr{g}_{t\varphi} =&\ \lambda^2(\delta^2 \mathscr{g}_{TT} - 2\gamma\delta\mathscr{g}_{T\phi} + \gamma^2\mathscr{g}_{\phi\phi})\,, \nonumber \\
\mathscr{g}_{\varphi\varphi} =&\ \lambda^2(\beta^2 \mathscr{g}_{TT} - 2\alpha\beta\mathscr{g}_{T\phi} + \alpha^2\mathscr{g}_{\phi\phi})\,, \nonumber \\
\mathscr{g}_{rr} =&\ \mathscr{g}_{\rho\rho}/(\partial_\rho r)^2\,, \nonumber \\
\mathscr{g}_{\vartheta\vartheta} =&\ \mathscr{g}_{\psi\psi}/(\partial_\psi \vartheta)^2\,, \nonumber 
\end{align}
It is only when $\alpha, \beta, \gamma$ and $\delta$ are constants that the new metric components \eqref{eq:New_BL_Metric} are independent of $t$ and $\varphi$.

It is clear that the equations \eqref{eq:DEs-2} admit $\alpha, \beta, \gamma$ and $\delta$ all being constants as a solution. 



\section{Technical Details}
\label{AppB:Technical-Details}

We collect the expressions for the various coefficient functions introduced above in Sec. \ref{sec:SecII-Matter-in-Doubly-Separable-Metrics}. We work in coordinates in which $R(r) = r$.


\subsection{Rigid Rotation, Sec. \ref{sec:SecIIB-Rigid-Rotation}}

The coefficient functions $\alpha_{0i}$ appearing in eq. \ref{eq:Coeff_Coeff} are 
\begin{align} \label{eq:alpha-0i}
\alpha_{00} =&\ -ar^4\left[2r^2g\partial_r^2f - r^2\partial_rg \cdot \partial_rf + 4g^2(1-f)\right]\,, \\
\alpha_{02} =&\ -2a^3r^2\left[2r^2g\partial_r^2f - (r\partial_rg- 4g)r\partial_rf 
\right. \nonumber \\ 
& \left. 
+ (r\partial_rg + 6g(1-g))(1-f)\right]\,, \nonumber \\
\alpha_{04} =&\ -a^5\left[2r^2g\partial_r^2f - (r\partial_rg- 8g)r\partial_rf  
\right. \nonumber \\
& \left. 
+ 2(r\partial_rg-2g)(1-f)\right]\,, \nonumber
\end{align}
from which we can see that for $f=1$, all of them vanish. The coefficient functions $\alpha_{1i}$ appearing in eq. \ref{eq:Coeff_Coeff} are 
\begin{widetext}
\begin{align} \label{eq:alpha-1i}
\alpha_{10} =&\ r^4\left[2r^2(r^2+2a^2)g\partial_r^2f - 
r^2(r^2+2a^2)\partial_rg\cdot\partial_rf 
+2(r^3\partial_rg - 4a^2g^2 - 2r^2g)f + 4(r^2+2a^2)g^2\right]\,, \\
\alpha_{12} =&\ a^2r^2\left[2r^2(r^2+4a^2)g\partial_r^2f - r(r(r^2+4a^2)\partial_rg - 8(r^2+2a^2)g)\partial_rf + 2(r(r^2 - 2a^2)\partial_rg + 12(r^2+a^2)g^2 - 2(5r^2+6a^2)g)f 
\right. \nonumber \\
&\ \left. + 4(r(r^2+a^2)\partial_rg - 3(r^2+2a^2)g^2 + 2(r^2+3a^2)g) \right]\,, \nonumber \\
\alpha_{14} =&\ a^4\left[-2r^2(r^2 - 2a^2)g\partial_r^2f + r(r(r^2-2a^2)\partial_rg + 16a^2g)\partial_rf + 2(r(r^2-2a^2)\partial_rg - 4r^2g^2 + 2(3r^2+2a^2)g)f \right. \nonumber \\
&\ \left. + 4(r(r^2+a^2)\partial_rg + 5r^2g^2) - 2 (3r^2 + a^2)g\right]\,, \nonumber \\
\alpha_{16} =&\ a^6\left[-2r^2g\partial_r^2f + (r^2\partial_rg -8rg)\partial_rf + 2(r\partial_rg - 2g)f + 4g^2\right]\,. \nonumber
\end{align}
\end{widetext}
The coefficient functions $\alpha_{2i}$ appearing in eq. \ref{eq:Coeff_Coeff} are 
\begin{widetext}
\begin{align} \label{eq:alpha-2i}
\alpha_{20} =&\ a\left[-2r^6(r^2+a^2)g\partial_r^2f + r^6(r^2+a^2)\partial_rg\cdot\partial_rf + 2r^4(r^3\partial_rg -2a^2g^2 - 2r^2g)(1-f)\right]\,, \\
\alpha_{22} =&\ ar^2\left[2r^2(r^2+a^2)(r^2-2a^2)g\partial_r^2f + r(r^2+a^2)(r(r^2-2a^2)\partial_rg + 8a^2g)\partial_rf \right. \nonumber \\
&\ \left. - 2(r(r^4-a^2r^2+a^4)\partial_rg - 6a^2(2r^2+a^2)g^2 - 2(r^4-5a^2r^2-3a^4)g)(1-f)\right]\,, \nonumber \\
\alpha_{24} =&\ a^3\left[2r^2(r^2+a^2)(2r^2-a^2)g\partial_r^2f - r(r(2r^4 + a^2r^2 - a^4)\partial_rg - 8(r^4 - a^4)g)\partial_rf \right. \nonumber \\
&\ \left. - 2(r(r^4-a^2r^2+a^4)\partial_rg + 2(5r^4+2a^2r^2)g^2 -2(5r^4 + 3a^2r^2 + a^4)g)(1-f)\right]\,, \nonumber \\
\alpha_{26} =&\ a^5\left[2r^2(r^2+a^2)g\partial_r^2f -r(r^2+a^2)(r\partial_rg - 8g)\partial_rf + 2(a^2r\partial_rg -2r^2g^2 - 2a^2g)(1-f)\right]\,, \nonumber
\end{align}
\end{widetext}
from which we also can see that for $f=1$, all of them vanish.


\subsection{Equal Tangential Pressures, Sec. \ref{sec:SecIIC-Tangential}}

The coefficient functions $\gamma_{2i}$ appearing in eq. \ref{eq:Quadratic-Quartic} are
\begin{widetext}
\begin{align} \label{eq:gamma-ij}
\gamma_{20} =&\ r^4\left[4r^4(1-g)g^2\partial_r^2f -2r^4(1-g)g\partial_rg\cdot\partial_rf + (r^4(\partial_rg)^2 -4r^2g^2(1-g^2))f + a^2r^2(\partial_rg)^2 + 4a^2r(1-g)g\partial_rg \right. \\
&\ \left. + 4(1-g)g^2((2r^2-a^2)g + a^2)\right]\,, \nonumber  \\
\gamma_{22} =&\ 2a^2r^2\left[4r^4(1-g)g^2\partial_r^2f - 2r^3(r\partial_rg - 4g)(1-g)g\partial_rf + r^2(r^2(\partial_r g)^2 - 4r(1-g)g\partial_r g - 12(1-g)^2g^2)f \right. \nonumber \\
&\ \left. + a^2r^2(\partial_r g)^2 + 4(2r^2 - a^2)(1-g)^2g^2\right]\,, \nonumber \\
\gamma_{24} =&\ a^4\left[4r^4(1-g)g^2\partial_r^2f - 2r^3(r\partial_rg - 8g)(1-g)g\partial_rf + r^2(r^2(\partial_rg)^2 - 8r(1-g)g\partial_rg + 4(3-g)(1-g)g^2)f \right. \nonumber \\
&\ \left. + a^2r^2(\partial_rg)^2 - 4a^2r(1-g)g\partial_rg - 4(1-g)g^2((2r^2+a^2)g - a^2)\right]\,, \nonumber
\end{align}
\end{widetext}
from which we can see that for $g=1$, all of them vanish.

The coefficient functions $\gamma_{4i}$ appearing in eq. \ref{eq:Quadratic-Quartic} are long and uninsightful (they do not reduce to simple values when $g=1$ or $f=1$). For example, 
\begin{widetext}
\begin{align}
\gamma_{40} =&\ r^6\left[4r^6g^2(\partial_r^2f)^2 + (-4r^6g\partial_rg\cdot\partial_rf + 8r^4(r\partial_rg - 2g)gf + 16a^2r^3g\partial_rg + 16r^2(r^2+2a^2)g^3 - 32a^2r^2g^2)\partial_r^2f +  \right. \nonumber \\
&\ \left. + r^6(\partial_rg)^2(\partial_rf)^2 - (4r^4(r\partial_rg-2g)\partial_rg \cdot f + 8a^2r^3(\partial_rg)^2 + (8r^2(r^2+2a^2)g^2 - 16a^2r^2g)\partial_rg)\partial_rf \right. \nonumber \\
&\ \left. + 4r^2(r\partial_rg - 2g)^2 f^2 + (16r(r^2-2a^2)g^2\partial_rg - 32g^2(2a^2g^2 + (r^2-2a^2)g))f + 32a^2 r g^2\partial_rg + 16(r^2+4a^2)g^4 - 64a^2g^3
\right]\,. \nonumber 
\end{align}
\end{widetext}
For this reason, we will refrain from showing the remaining $\gamma_{4i}$ coefficients and direct the reader to the accompanying $\mathtt{Mathematica}$ notebooks \cite{Kocherlakota+2025}.


\subsection{Equal Meridional Pressures, Sec. \ref{sec:SecIID-Meridional}}

The coefficient functions $p_{\vartheta i}$ appearing in eq. \ref{eq:Meridional-Pressures} are
\begin{align} \label{eq:pth-ij}
p_{\vartheta 0} =&\ r^5\left[2rg\partial_r^2f - (r\partial_rg - 4g)\partial_rf -2\partial_rg f\right]\,, \\
p_{\vartheta 2} =&\ 4a^2r^4g\partial_r^2f - 2a^2r^3(r\partial_rg - 6g)\partial_rf \nonumber \\ 
&\ - 4a^2r^2(r\partial_rg - g^2)f + 4a^2r^2(1-2g)g\,, \nonumber \\
p_{\vartheta 4} =&\ 2a^4r^2g\partial_r^2f - a^4r(r\partial_rg - 8g)\partial_rf \nonumber \\
&\ - 2a^4(r\partial_rg - 2g)f - 4a^4g^2\,. \nonumber 
\end{align}
The coefficient functions $p_{r i}$ appearing in eq. \ref{eq:Meridional-Pressures} are
\begin{align} \label{eq:pr-ij}
p_{r 0} =&\ 4r^5g\partial_rf + 4r^4gf - 4r^4g^2 \,, \\
p_{r 2} =&\ 4a^2r^3g\partial_rf + 4a^2r^2(2-g)gf - 4a^2r^2g\,. \nonumber
\end{align}


\end{appendix}
\end{document}